\documentclass[a4paper,accepted=2019-07-30]{quantumarticle}

\pdfoutput=1

\usepackage[a4paper,left=2cm,right=2cm,top=3cm,bottom=3cm]{geometry}
\usepackage{soul}
\usepackage{relsize}
\usepackage{slantsc}
\usepackage{graphicx}
\usepackage{amsmath}
\usepackage{amssymb}
\usepackage{amsthm}
\usepackage{amsbsy}
\usepackage{mathrsfs}
\usepackage{varioref}
\usepackage{dsfont}
\usepackage{bm}
\usepackage{color}
\usepackage[usenames,dvipsnames]{xcolor}
\usepackage[colorlinks = false, citecolor=green, linkcolor=red, breaklinks=true, linktocpage=true]{hyperref}
\usepackage{cleveref}
\usepackage[font = footnotesize, labelfont = bf, justification = RaggedRight]{caption}
\usepackage{array} 
\usepackage{paralist} 
\usepackage{verbatim} 
\usepackage[parfill]{parskip}
\usepackage[numbers, sort&compress]{natbib}

\newcommand{\sub}[1]{_{\!\mathsmaller{\, #1}}}
\newcommand{\eq}[1]{Eq.~\eqref{#1}}
\newcommand{\fig}[1]{Fig.~\ref{#1}}

\newcommand{\app}[1]{Appendix~(\ref{#1})}

\newcommand{\corref}[1]{Corollary~\ref{#1}}

\newcommand{\reqref}[1]{Requirement~\ref{#1}}
\newcommand{\ts}{\textsuperscript}
\newcommand{\<}{\langle}
\renewcommand{\>}{\rangle}
\newcommand{\ket}[1]{|{#1}\rangle}

\newcommand{\prs}[1]{P\sub{\s}[{#1}]}
\newcommand{\pra}[1]{P\sub{\aa}[{#1}]}

\newcommand{\h}{{\mathcal{H}}}
\newcommand{\w}{{\mathcal{W}}}

\newcommand{\s}{{\mathcal{S}}}
\newcommand{\e}{{\mathcal{E}}}
\newcommand{\xx}{{\mathcal{X}}}

\renewcommand{\aa}{{\mathcal{A}}}
\newcommand{\bb}{{\mathcal{B}}}

\newcommand{\mm}{{\mathcal{M}}}

\newcommand{\ii}{{\mathcal{I}}}
\newcommand{\jj}{{\mathcal{J}}}
\newcommand{\co}{\mathds{C}}

\newcommand{\one}{\mathds{1}}

\newcommand{\tr}{\mathrm{tr}}

\renewcommand{\ln}[1]{\mathrm{ln} \left( {#1}\right)}

\theoremstyle{plain}

\newtheorem{req}{Requirement}

\begin{document}

\title{Conditional work statistics of quantum measurements}

\author{M. Hamed Mohammady}
\affiliation{Department of Physics, Lancaster University, LA1 4YB, United Kingdom}
\affiliation{RCQI, Institute of Physics, Slovak Academy of Sciences, D\'ubravsk\'a cesta 9, Bratislava 84511, Slovakia}

\author{Alessandro Romito}
\affiliation{Department of Physics, Lancaster University, LA1 4YB, United Kingdom}


\begin{abstract}
In this paper we introduce a definition for conditional energy changes due to general quantum measurements, as the change in the conditional energy  evaluated before, and after, the measurement process. By imposing minimal physical requirements on these conditional energies, we show that the most general expression for the conditional energy after the measurement is simply the expected value of the Hamiltonian given the post-measurement state. Conversely, the conditional energy before the measurement process is shown to be given by the real component of the weak value of the Hamiltonian.  Our definition generalises well-known notions of distributions of internal energy change, such as that given by stochastic thermodynamics. By determining the conditional energy change of both system and measurement apparatus, we obtain the full conditional work statistics of quantum measurements, and show that this vanishes for all measurement outcomes if the measurement process conserves the total energy. Additionally, by incorporating the measurement process within a cyclic heat engine, we quantify the non-recoverable work due to measurements. This is shown to always be non-negative, thus satisfying the second law, and will be independent of the apparatus specifics for two classes of projective measurements.  

\end{abstract}

\maketitle

\section{Introduction}

Measurements play an important role in thermodynamic processes. This has been established ever since the introduction of Maxwell's  demon \cite{Maxwell1871} and the subsequent insights gained in the thermodynamic role of information \cite{Szilard1976,Landauer1961,Penrose1970,Landauer1996,Bennett2003,Maruyama2009}. 
In the quantum regime, measurements are even more intimately linked to thermodynamics
\cite{Goold2015,Vinjanampathy2016,Millen2016}. On the one hand, energy measurements are essential to  extend the laws of thermodynamics in the form of fluctuation theorems  \cite{Campisi2011,Funo2013,Allahverdyan2014,Miller2016,Aberg2016,Perarnau-Llobet2016a,Lostaglio2018}.  On the other hand, measurement processes typically involve the exchange of energy  between a system and  detector, and the fundamental energy cost of quantum measurements is a subject of intense study \cite{Sagawa2009b,Jacobs2012a,Navascues2014a,Abdelkhalek2016a,Guryanova2018}.

In quantum mechanics, measurements induce an unavoidable  stochastic change in the state of a system, which will generally  modify  its energy.  How the energy will change on average is well understood, and is simply given by the difference in the system's average energy, evaluated before and after the measurement. 
 Quantifying the change in energy, \emph{conditional} on observing a given measurement outcome, however, is still lacking a general answer. A well known method used to establish the energetic fluctuations due to dynamical processes, such as measurement, is the Two-Point-Measurement (TPM) protocol, which uses projective energy measurements before, and after, the dynamical process in question  \cite{Campisi2011}. This protocol, however, is known to break down when the system initially has coherences with respect to its Hamiltonian \cite{Allahverdyan2014, Lostaglio2018}.  An alternative approach is that used in quantum stochastic thermodynamics \cite{Horowitz2012,Hekking2013, Alonso2016, Alexia-measurement-thermodynamics, Naghiloo2017, Naghiloo2018, Elouard-Book}, wherein the system follows a trajectory of pure states that are not necessarily eigenstates of the Hamiltonian. The change in energy is thus defined as the difference in expected values of the Hamiltonian at the start and end of the trajectory in question.  Such an approach, however, implicitly assumes that we know which pure state the system initially occupies.

In the present paper we provide a general definition for conditional energy changes, for  general quantum measurements and initial system states, as the difference in conditional energies evaluated before and after the measurement process. By imposing three minimal physical requirements on these conditional energies, we show that the most general expresion for the conditional energy after the measurement is simply the expected value of the Hamiltonian given the post-measurement state. Conversely, the conditional energy before the measurement process is shown to be given by the real component of the generalised weak value of the Hamiltonian \cite{Aharonov1988, Haapasalo2011,Romito2010, Dressel2014, Hovhannisyan2019a}.  
The energetic statistics obtained by the proposed definition  generalises existing results in the literature, which are valid in specific circumstances:  (i) if the measured observable involves an initial and final energy measurement, we regain the work distribution of the TPM protocol; (ii)  if the observable measured is the Heisenberg-evolved Hamiltonian, the energy statistics is equivalent to the quasi-probability distribution over the random variable of work introduced in \cite{Allahverdyan2014}; and  (iii) if the measurement process first projects the system onto one of its pure state components, we obtain the definition for internal energy change along a quantum trajectory used in stochastic thermodynamics. 

By  evaluating the conditional energy change of both system and measurement apparatus, we obtain the full conditional work statistics of quantum measurements. We show that when the measurement process conserves the total energy, the conditional work vanishes for all measurement outcomes, and not just on average.   Finally, by incorporating the measurement process within a cyclic heat engine involving a single heat bath of temperature $T$, we define the non-recoverable, or irreversible, work due to measurement. This is shown to be non-negative, thus  satisfying the second law of thermodynamics. In general, the non-recoverable work will depend on the specifics of the measurement apparatus. However, we show that it becomes a system-only property for two classes of projective measurements: (i) repeatable projective measurements, which is a generalisation of ideal projective measurements (the case for ideal projective measurements has already been shown in \cite{Abdelkhalek2016a});  and (ii)  ``noisy'' projective measurements where the apparatus has the same dimension as the system. Here, a noisy projective measurement is to be understood as the necessarily non-repeatable measurement of an observable by use of a measurement apparatus that is initially prepared in a state of full-rank.

\section{General Measurements}

\begin{figure}[!htb]
 \includegraphics[width = 0.45\textwidth]{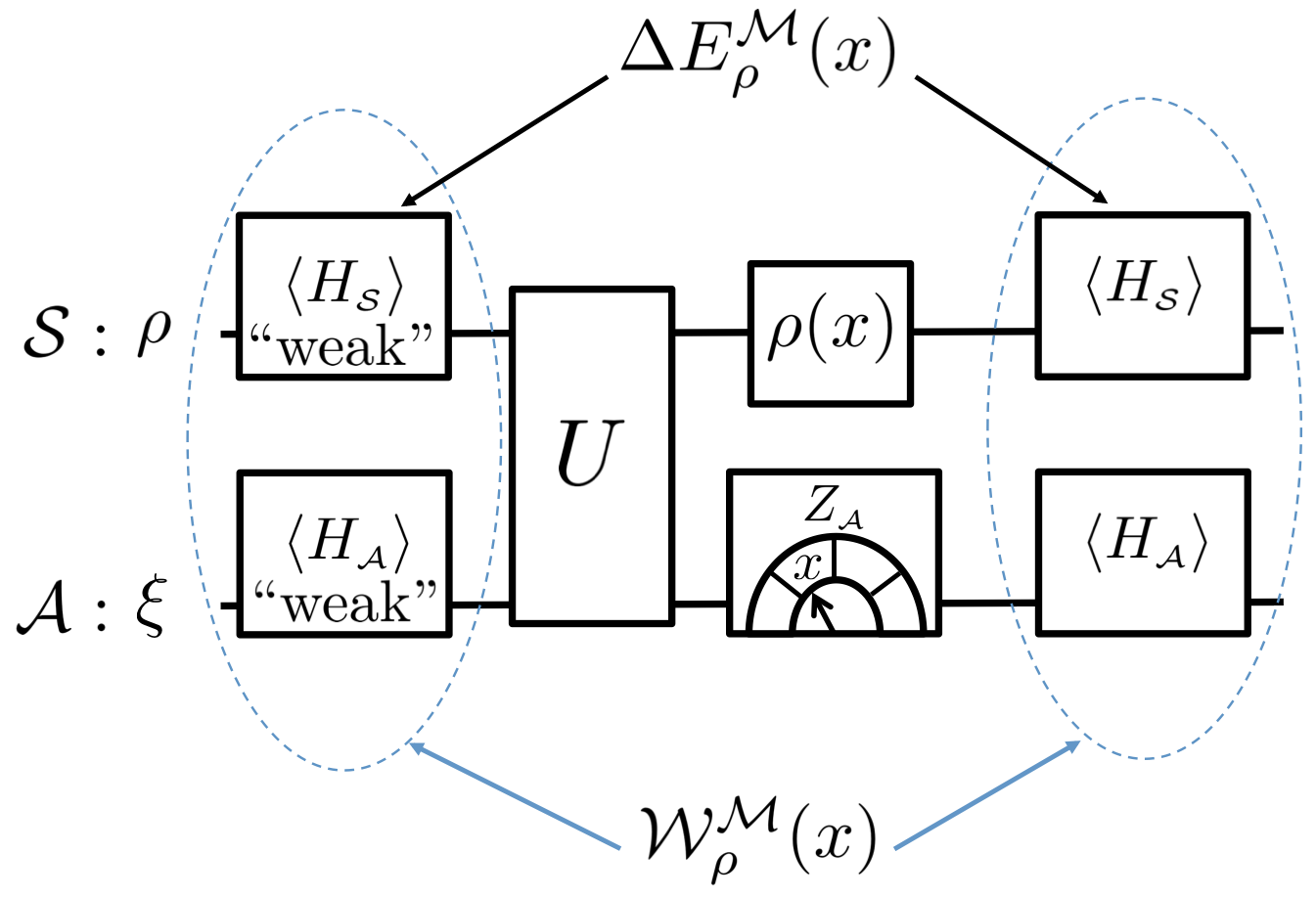} 
\caption{Conditional energy changes due to a measurement model $\mm$. Initially, the  energy of system $\s$ and apparatus $\aa$ are measured ``weakly''. Subsequently,  $\s$ is measured by the POVM $M$ via a unitary coupling $U$ with $\aa$, followed by a projective measurement of $\aa$ by   $Z\sub{\aa}$.  For each outcome $x$ of $M$, the system is transformed to $\rho(x)$. At the end of the measurement process, the average energy of both $\s$ and $\aa$ are measured. The  conditional energy change of $\s$ is $\Delta E_\rho^\mm(x)$, while that of the compound system is the conditional work $\w_\rho^\mm(x)$. }\label{fig:Measurement}
\end{figure}

Observables on a quantum system $\s$, with Hilbert space $\h\sub{\s}$, are described by positive operator valued measures (POVMs) $M := \{M_x\}_{x \in \xx}$. Here $\xx$ denotes the outcome set (readouts) of the measurement, and $M_x$ are  positive operators (referred to as effects or POVM elements) acting on $\h\sub{\s}$ that sum to the identity, and give the probability of observing outcome $x$ via the Born rule \cite{PaulBuschMarianGrabowski1995, Busch1996, Busch2016a}. The POVM description of measurements, however, is insufficient for energetic considerations.  Measurement, seen as a  \emph{physical process}, is implemented by coupling  the system  to a detecting apparatus, which is then subjected to projective measurements. Indeed, each POVM $M$ admits infinitely many physical implementations, or measurement models/processes, described by the tuple $\mm := (\h\sub{\aa}, \xi, U, Z\sub{\aa})$ \cite{Busch2016a}. Here $\h\sub{\aa}$ is the Hilbert space of apparatus $\aa$, and $\xi$ is the state in which the apparatus is initially prepared; $U$ is a ``premeasurement'' unitary operator acting on $\h\sub{\s}\otimes \h\sub{\aa}$ due to turning on an interaction between $\s$ and $\aa$ at time $t=0^+$ and turning it off again at $t=\tau^-$; and $Z\sub{\aa} = \sum_{x\in \xx} x P\sub{\aa}^x$ is a self-adjoint operator defining a sharp observable, or projective valued measure (PVM), which  $\aa$ is measured by after the premeasurement interaction with $\s$, i.e. at time $t=\tau$, such that each outcome $x$ of $Z\sub{\aa}$ is associated with the same for the POVM $M$ on $\s$.  The measurement model is depicted in \fig{fig:Measurement}. 

For each measurement outcome $x$, the measurement model defines an instrument \cite{Heinosaari2011} on $\s$, given as
\begin{align}\label{eq:instrument-outcome}
\ii_x ^\mm (\rho) := \tr\sub{\aa}[(\one\sub{\s}\otimes P\sub{\aa}^x) U(\rho \otimes \xi) U^\dagger]. 
\end{align}
The instrument describes how the state of the system changes due to measurement. An ideal measurement of a PVM $M$ is implemented by a L\"uders instrument  $\ii_x^\mm(\rho) = M_x \rho M_x$.

At the end of the measurement process, the compound system $\s+\aa$ will be in the state 
\begin{align}\label{eq:system-apparatus-after-measurement}
\varrho\sub{\s+\aa} &:= \sum_{x\in \xx} P\sub{\aa}^x U(\rho\otimes \xi)U^\dagger P\sub{\aa}^x, \nonumber \\
& = \sum_{x \in \xx} p_\rho^M(x) \varrho\sub{\s+\aa}(x),
\end{align}
where $p_{\rho}^M (x) := \tr[M_x \rho] \equiv \tr[\ii_x^\mm(\rho)]$ is the probability of observing outcome $x$ of $M$, given the initial system state $\rho$, and  $\varrho\sub{\s+\aa}(x) := P\sub{\aa}^x U(\rho\otimes \xi)U^\dagger P\sub{\aa}^x/ p_\rho^M(x)$. Since the  states $\varrho\sub{\s+\aa}(x)$ are orthogonal with respect to the Hilbert-Schmidt inner product, $\varrho\sub{\s+\aa}$ is a statistical ensemble, or ``Gemenge'', $\{p_\rho^M(x), \varrho\sub{\s+\aa}(x)\}$ and thus  offers an ignorance interpretation. As a result, the ideal projective measurement of $\aa$ by the observable $Z\sub{\aa}$, after premeasurement, is often referred to as the ``objectification'' process \cite{PeterMittelstaedt2004}. Finally, we shall refer to the conditional states of $\s$ after observing the measurement outcome $x$ as  $\rho(x):= \ii_x^\mm(\rho)/p_{\rho}^M (x) \equiv \tr\sub{\aa}[\varrho\sub{\s+\aa}(x)]$, and similarly the orthogonal states of $\aa$ representing outcome $x$ will be denoted $\xi(x):=  P\sub{\aa}^x \xi'  P\sub{\aa}^x / p_{\rho}^M (x) \equiv \tr\sub{\s}[\varrho\sub{\s+\aa}(x)]$, where $\xi':= \tr\sub{\s}[U(\rho \otimes \xi)U^\dagger]$ is the state of $\aa$ after premeasurement.

\section{Energy change conditional on measurement outcome}

We wish to quantify the increase in energy of the system, conditional on observing outcome $x$ of the POVM $M$. Before doing so, it will be instructive to consider the same question for a classical probabilistic system. Consider the time-dependent and time-independent random variables $A(t)$ and $B$, respectively, with the joint probability distributions $p(a(t), b)$ such that $\sum_{a(t)} p(a(t), b) = p(b)$ for all $t$. The change in the expected value of $A(t)$, for $t=0$ and $t=\tau$, conditional on $B=b$, can be defined as 
\begin{align}\label{eq:classical-conditional-exp-change}
\Delta \<A(t)\>_b &:= \< A(\tau)\>_b - \<A(0)\>_b.
\end{align}
Here, the expected value of $A(t)$ conditional on $B=b$ is defined as $\<A(t)\>_b := \sum_{a(t)} a(t) \, p(a(t)|b)$, where $p(a(t)|b) :=  p(a(t),b)/p(b) = p(a(t),b)/\sum_{a(t)} p(a(t),b)$ is the probability of $A(t)=a(t)$ conditional on $B=b$.  Clearly, averaging $\Delta \<A(t)\>_b$ with respect to the probability distribution $p(b)$ yields $\sum_b p(b) \Delta \<A(t)\> = \sum_{a(\tau)} p(a(\tau))a(\tau) - \sum_{a(0)} p(a(0))a(0) \equiv \<A(\tau)\> - \<A(0)\>$. If $A(t)$ is the energy of the system, while $B$ is another physical quantity, then \eq{eq:classical-conditional-exp-change} will quantify the conditional increase in energy for classical systems.

Now consider a quantum measurement process $\mm$ which takes place during the time interval $(0, \tau)$. We wish to quantify the increase in energy of the system, conditional on observing outcome $x$ of the POVM $M$. To this end, we may define the conditional energies $E_{x, t}^\mm(\rho)$ for $t=0$ and $t=\tau$, from which the conditional increase in energy will be given as 

\begin{align}\label{eq:conditional-energy-change}
\Delta E_\rho^\mm(x) & := E_{x, \tau}^\mm(\rho) - E_{x, 0}^\mm(\rho).
\end{align} 

\eq{eq:conditional-energy-change} can be seen as a quantum  analogue to \eq{eq:classical-conditional-exp-change}. Here, we  replace $A(t)$ with the time-dependent Hamiltonian $H\sub{\s}(t) := \sum_j \epsilon_j(t) P\sub{\s}^j(t)$, where $\epsilon_j(t)$ are the energy-eigenvalues and $P\sub{\s}^j(t)$ the corresponding spectral projections. Similarly, $B$ is replaced with the POVM $M$. For the two equations to be fully analogous, however, we must obtain a quantum version of the joint probability distribution $p(\epsilon_j(t), x)$. Prima facie, this would be determined by the Born rule, given the initial state $\rho$ and the joint measurement of $H\sub{\s}(t)$ and $M$.   However, due to contextuality, $p(\epsilon_j(t), x)$ is generally not uniquely defined, and will depend on the \emph{order} in which we measure $H\sub{\s}(t)$ and $M$: if we first perform an ideal measurement of $H\sub{\s}(t)$, and then measure $M$, we obtain $p(\epsilon_j(t), x) = \tr[\ii_x^\mm(P\sub{\s}^j(t) \rho P\sub{\s}^j(t))]$, whereas if we first measure $M$, and then $H\sub{\s}(t)$, we obtain $p(\epsilon_j(t), x) = \tr[P\sub{\s}^j(t) \ii_x^\mm(\rho)]$.  Note that these joint probability distributions may differ even if $\rho$ is a classical probabilistic mixture of energy eigenstates -- it is the quantum nature of measurements that leads to the contextuality of the joint probability distribution. We define the ``classical'' (non-contextual) limit as being characterised by measurement processes $\mm$ which satisfy  $\tr[\ii_x^\mm(P\sub{\s}^j(t) \rho P\sub{\s}^j(t))] = \tr[P\sub{\s}^j(t) \ii_x^\mm(\rho) ]$ for all $\rho, j, x$, thereby providing a uniquely defined $p(\epsilon_j(t), x)$, so that  \eq{eq:conditional-energy-change} reduces to \eq{eq:classical-conditional-exp-change}. 

We are therefore interested in obtaining the most general expressions of $E_{x, t}^\mm(\rho)$, for $t=0$ and $t= \tau$, that are valid for all states $\rho$ and POVMs $M$, so that \eq{eq:conditional-energy-change} will be a physically meaningful definition for conditional energy change which, in the ``classical'' limit,  will reduce to \eq{eq:classical-conditional-exp-change}. To this end, we impose the following three minimal requirements on  $E_{x, t}^\mm(\rho)$:

\begin{req}\label{requirement1}
For all $\rho$ and $\mm$, 
\begin{align}\label{eq:req1-avg-cond}
\sum_{x \in \xx} p_\rho^M(x) E_{x, t}^\mm(\rho) = \tr[H\sub{\s}(t) \rho_t],
\end{align}
where $\rho_0 := \rho$, and $\rho_\tau := \sum_{x\in \xx} \ii_x^\mm(\rho)$.
\end{req}
\begin{req}\label{requirement2}
For all $\rho$ and $\mm$, and all ensembles $\{p_k,\rho^{(k)}\}$ satisfying $\sum_k p_k \rho^{(k)} = \rho$,
\begin{align}\label{eq:cond-E-convex}
\sum_k \frac{p_k p_{\rho^{(k)}}^M(x)}{p_\rho^M(x)} E_{x, t}^\mm(\rho^{(k)}) = E_{x, t}^\mm(\rho),
\end{align}
where $\sum_k p_k p_{\rho^{(k)}}^M(x) = p_\rho^M(x)$.
\end{req}
\begin{req}\label{requirement3}
If  
\begin{align}\label{eq:classical-limit-condition}
\tr[\ii_x^\mm(P\sub{\s}^j(t) \rho  P\sub{\s}^j(t))] = \tr[P\sub{\s}^j(t) \ii_x^\mm(\rho) ]
\end{align}
for all $\rho$ and $j$, then
\begin{align}
E_{x, t}^\mm(\rho) &= \sum_j \epsilon_j(t) \frac{p(\epsilon_j(t) , x)}{p(x)}, \nonumber \\
&= \sum_j \epsilon_j(t) \frac{\tr[P\sub{\s}^j(t) \ii_x^\mm(\rho)]}{\sum_j \tr[P\sub{\s}^j(t) \ii_x^\mm(\rho)]}, \nonumber \\
& = \frac{\tr[H\sub{\s}(t) \ii_x^\mm(\rho)]}{\tr[\ii_x^\mm(\rho)]} =: \tr[H\sub{\s}(t) \rho(x)].
\end{align}
\end{req}
\reqref{requirement1} is necessary for the average conditional increase in energy to equal the increase in average energy, i.e. $\sum_{x\in \xx} p_\rho^M(x) \Delta E_\rho^\mm(x) = \tr[H\sub{\s}(\tau)\rho_\tau] - \tr[ H\sub{\s}(0) \rho]$;   \reqref{requirement2} is needed to ensure that the conditional increase in energy  is not dependent on how the state $\rho$ is  prepared; and \reqref{requirement3} follows from the fact that in the ``classical'' limit, achieved when \eq{eq:classical-limit-condition} is satisfied, the conditional energies $E_{x,t}^\mm(\rho)$ will be evaluated by sampling the   energy eigenvalues by the conditional probability distribution $p(\epsilon_j(t) | x) := p(\epsilon_j(t), x)/p(x) = \tr[P\sub{\s}^j(t) \ii_x^\mm(\rho)]/\tr[ \ii_x^\mm(\rho)]$.

Following the logic of conditional expectation values in classical probability theory, we may be tempted to define the conditional energy before the measurement, $E_{x, 0}^\mm(\rho)$, as
\begin{align}\label{eq:conditional-exp-before}
\<H\sub{\s}(0)\>_{(\rho,x)}:= \sum_j \epsilon_j(0) \frac{\tr[\ii_x^\mm(P\sub{\s}^j(0) \rho P\sub{\s}^j(0))]}{\sum_j \tr[\ii_x^\mm(P\sub{\s}^j(0) \rho  P\sub{\s}^j(0))]}, 
\end{align}
and the conditional energy after the measurement, $E_{x, \tau}^\mm(\rho)$, as
\begin{align}\label{eq:conditional-exp-after}
\<H\sub{\s}(\tau)\>_{(\rho,x)} &:= \sum_j \epsilon_j(\tau) \frac{\tr[P\sub{\s}^j(\tau)\ii_x^\mm( \rho) ]}{\sum_j \tr[P\sub{\s}^j(\tau) \ii_x^\mm(\rho)]}, \nonumber \\
& = \tr[H\sub{\s}(\tau) \rho(x)].
\end{align}
Here, \eq{eq:conditional-exp-before} obtains the joint probabilities $p(\epsilon_j(0), x)$  by  performing an ideal measurement of $H\sub{\s}(0)$ at $t=0$, i.e. before the measurement process $\mm$. In contrast, \eq{eq:conditional-exp-after} obtains $p(\epsilon_j(\tau), x)$ by performing an ideal measurement of $H\sub{\s}(\tau)$ at $t= \tau$, i.e., after the measurement process $\mm$. These equations trivially satisfy \reqref{requirement3}. Moreover, as shown in \app{appendix:proof-of-weak-value}, \eq{eq:conditional-exp-after} is the most general expression for $E_{x,\tau}(\rho)$ that will satisfy all three requirements. Consequently, we may safely choose
 \begin{equation}
 \label{eq:energy-after}
 E_{x,\tau}^\mm(\rho) := \tr[H\sub{\s}(\tau) \rho(x)].
 \end{equation}
However, \eq{eq:conditional-exp-before} does not satisfy \reqref{requirement1} and \reqref{requirement2}, as \eq{eq:req1-avg-cond} and \eq{eq:cond-E-convex} are satisfied only if either $\rho$ or $M_x$ commute with $H\sub{\s}(0)$. In Appendix \ref{appendix:proof-of-weak-value} we prove that the most general form of $E_{x,0}^\mm(\rho)$ that is compatible with all three physical requirements  is
 \begin{align}\label{eq:general-form-f}
E_{x,0}^\mm(\rho) = \lambda \, {}_x\langle H\sub{\s}(0) \rangle_\rho + (1-\lambda) \, {}_x\langle H\sub{\s}(0) \rangle_\rho^* 
 \end{align} 
 where $\lambda \in [0,1]$ and  
 \begin{align}\label{eq:weak-value}
 {}_x\langle H\sub{\s}(0) \rangle_\rho := \frac{\tr[M_x H\sub{\s}(0) \rho]}{p_\rho^M(x)} \equiv \frac{\tr[\ii_x^\mm(H\sub{\s}(0) \rho)]}{p_\rho^M(x)}
 \end{align}
  is the generalised weak value of the Hamiltonian $H\sub{\s}(0)$, given the initial state $\rho$, and postselected by outcome $x$ of the POVM $M$ \cite{Aharonov1988, Haapasalo2011,Romito2010, Dressel2014}.
 
 The weak value ${}_x\langle H\sub{\s}(0) \rangle_\rho$ is generically a complex number associated with  conditional observables in quantum mechanics.
 While its imaginary part is a non-universal feature usually associated with the dynamics and back-action of the measurement process \cite{PhysRevA.85.012107}, the real part is the universal (independent of the particular measurement implementation) response of a detector in the limit of a vanishing measurement disturbance, and can be associated with the physical estimation of conditional quantities \cite{Steinberg1995,Romito2014}; indeed, as discussed in  \cite{PhysRevLett.104.240401}, the real component of ${}_x\langle H\sub{\s}(0) \rangle_\rho$ can be understood as the limit of \eq{eq:conditional-exp-before} as the disturbance of  $\rho$ due to the initial measurement of $H\sub{\s}(0)$ becomes vanishingly small.

 Therefore, \eq{eq:general-form-f} has an operationally meaningful interpretation when  $\lambda=1/2$, in which case it reduces to the real part of the weak value, and we shall use it as the definition of the initial conditional energy in \eq{eq:conditional-energy-change}, i.e.
 \begin{align}\label{eq:cond-E-weak}
 E_{x,0}^\mm(\rho) &:=\mathrm{Re}\left({}_x\langle H\sub{\s}(0) \rangle_\rho \right)\nonumber \\
 &\equiv \frac{\tr[(M_x H\sub{\s}(0) + H\sub{\s}(0) M_x) \rho]}{2 p_\rho^M(x)}.
 \end{align}

 We note that in the stochastic thermodynamics literature \cite{Horowitz2012,Hekking2013, Alonso2016, Alexia-measurement-thermodynamics, Naghiloo2017, Naghiloo2018, Elouard-Book}, the conditional change in energy along ``trajectories'', which can be thought of as being determined by outcomes of a general measurement, is defined as
 \begin{align}\label{eq:naiive-defn-energy-change}
\Delta \tilde{E}_\rho^\mm(x) := \tr[H\sub{\s}(\tau) \rho(x)] - \tr[H\sub{\s}(0) \rho].
\end{align} 
Here, the first term coincides with our definition for the conditional energy after the measurement process, i.e. $E_{x, \tau}^\mm(\rho)$. However, the second term is the average \emph{unconditional} energy of the system prior to measurement. While  \eq{eq:naiive-defn-energy-change} satisfies \reqref{requirement1}, it will fail to satisfy \reqref{requirement2}, since \eq{eq:cond-E-convex} will only be satisfied if $\rho$ is a pure state. Moreover, \eq{eq:naiive-defn-energy-change} clearly fails to satisfy  \reqref{requirement3}.

In \app{Conditional change in energy given multiple measurements} we show how \eq{eq:conditional-energy-change} can be generalised so as to give an additive value for conditional energy change due to any measurement $\mm_i$ from the sequence of measurements $\mm := \{\mm_i\}_{i=1}^I$, with outcomes $x := \{x_i\}_{i=1}^I$. This is done by introducing a trivial measurement $\mm_{I+1}$, with only one outcome, at the end of the sequence $\mm$, so that $\tr[\ii_x^{\mm_{I+1}}(A)]= \tr[A]$ for any operator $A$. The conditional increase in energy due to the $i\ts{th}$ measurement in the sequence can therefore be defined as $\Delta E_\rho^{\mm_i}(x) = E_{x_{i+1}, t_{i+1}}^{\mm_{i+1}}(\rho) - E_{x_{i}, t_{i}}^{\mm_{i}}(\rho)$, where
 \begin{align}
E_{x_{i}, t_{i}}^{\mm_{i}}(\rho):= \frac{\mathrm{Re}\left(\tr[\ii_x^\mm(i) (H\sub{\s}(t_i) \rho(x_{i-1})) ]\right)}{\tr[\ii_x^\mm(i) ( \rho(x_{i-1}))]}.
\end{align} 
Here $\ii_x^\mm(i) := \ii_{x_{I+1}}^{\mm_{I+1}}\circ \dots \circ \ii_{x_{i}}^{\mm_{i}}$ is the instrument implemented by the measurements $\mm_i$ through to $\mm_{I+1}$; $H\sub{\s}(t_i)$ is the system Hamiltonian before the measurement of $\mm_i$ in the sequence $\mm$; and $\rho(x_{i}) = \ii_{x_{i}}^{\mm_{i}}\circ \dots \circ \ii_{x_{1}}^{\mm_{1}} (\rho) / \tr[\mathrm{idem}]$ is the state of the system after the $i$\ts{th} measurement in the sequence. Here we denote $\rho(x_0) := \rho$.   It follows that $\sum_{i=1}^I \Delta E_\rho^{\mm_i}(x) = \Delta E_\rho^\mm(x)$, in full agreement with the conditional energy increase when the full sequence of measurements are considered.

\subsection{Relation between the proposed definition and previous work}
 The proposed definition of conditional energy change is rather general, and it encompasses and generalises known protocols. First, if a quantum system is  measured by the instruments $\ii_{(m,n)}^\mm(\rho):=  P\sub{\s}^n(\tau) U P\sub{\s}^m(0) \rho  P\sub{\s}^m(0) U^\dagger P\sub{\s}^n(\tau) $, where $U$ is a unitary operator while $P\sub{\s}^m(0)$ and $P\sub{\s}^n(\tau)$ are  the spectral projections of Hamiltonians $H\sub{\s}(0)$ and $H\sub{\s}(\tau)$, respectively,  then we obtain the probability distribution $p_\rho^M(m,n)$  over the random variable of work, $\epsilon_n(\tau) - \epsilon_m(0)$, given by the TPM protocol for an isolated quantum system that unitarily evolves by $U$ (see \app{appendix:Connection to the TPM protocol}). Secondly,  if the system is subject to an ideal measurement of the Heisenberg evolved Hamiltonian $U^\dagger H\sub{\s}(\tau) U$, the resulting probability distribution $p_\rho^M(n)$ over the general energy difference $\Delta E_\rho^\mm(n)$ will be equivalent to the  quasi-probability distribution  $\tilde p_{(m,n)}=\mathrm{Re}(\tr[ U^\dagger P\sub{\s}^n(\tau) U P\sub{\s}^m(0) \rho])$,  introduced in \cite{Allahverdyan2014}, over the random variable  $\epsilon_n(\tau) - \epsilon_m(0)$ (see \app{appendix:Connection to quasi-probability distributions of work}). 
 Finally, if $\rho$ is a mixture of  pure states $\ket{\psi_m}$, and the POVM $M$ defines a sequence of measurements with outcomes $x = (m,n,...)$, with the first outcome $m$ being due to a projective measurement with respect to the orthonormal basis $\ket{\psi_m}$, then the conditional initial energy of $\rho$ will be $E_{x,0}^\mm(\rho) =\<\psi_m| H\sub{\s} (0)|\psi_m\>$  (see \app{appendix:Connection to quantum stochastic thermodynamics}). This coincides with the definition of initial internal energy of a system along a quantum trajectory \cite{Horowitz2012,Hekking2013, Alonso2016, Alexia-measurement-thermodynamics, Naghiloo2017, Naghiloo2018, Elouard-Book}.
 
\begin{figure}[t]
\includegraphics[width = 0.45\textwidth]{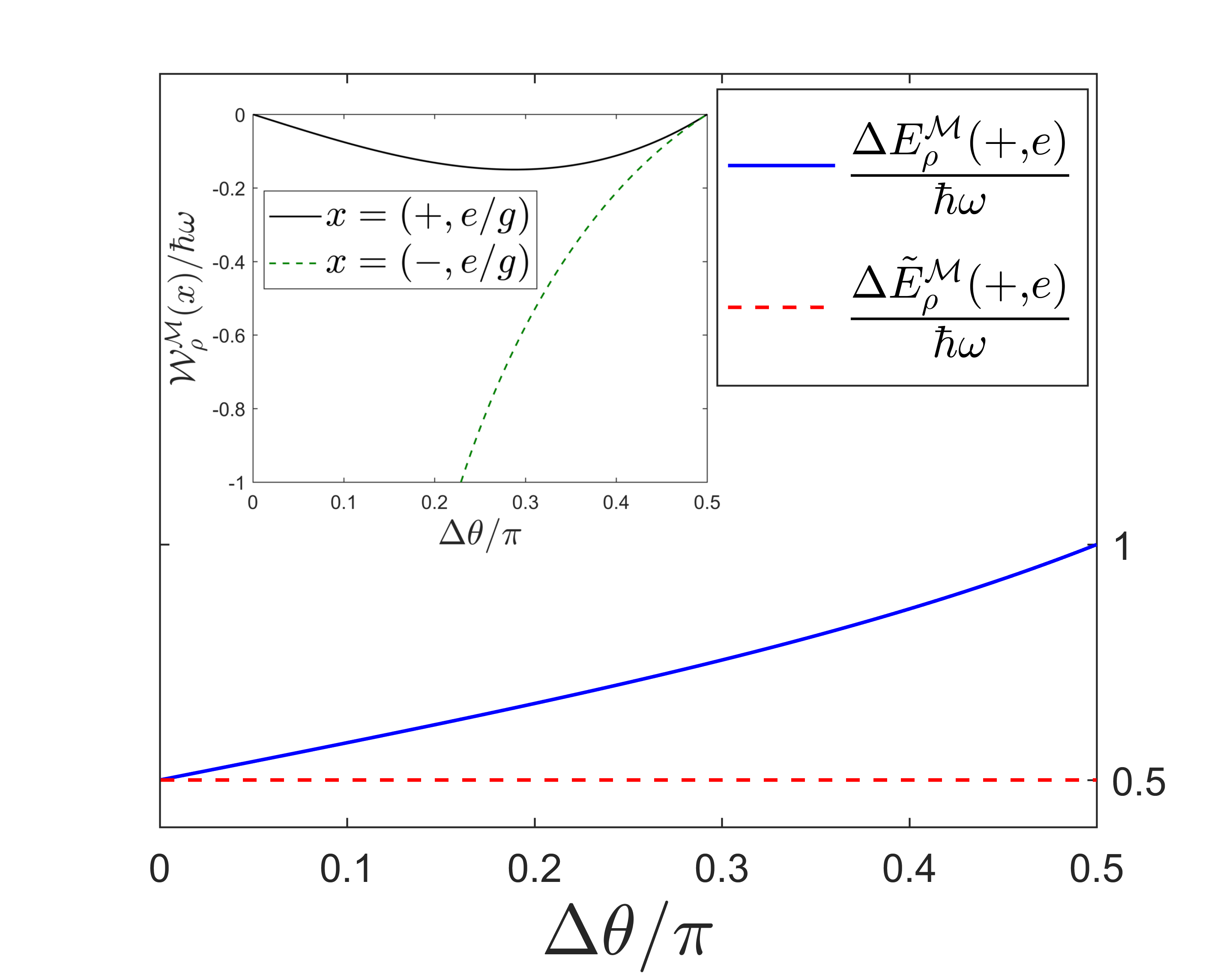} 
\caption{Conditional change in energy of a qubit due to a sequence of measurements $\mm := \{\mm_1, \mm_2\}$. The system is initially prepared in a pure state $\ket{ \pi/2 ,+}$. The first observable $\mm_1$ projects the system onto states $\ket{\theta_2,\pm}$, while the measurement of the observable $\mm_2$ takes it to states $\ket{e/g}$.  The conditional change of energy $\Delta E_\rho^\mm(x)$ can be computed by \eq{eq:conditional-energy-change}. Alternatively, the unconditional change in energy $\Delta \tilde E_\rho^\mm(x)$ may be computed by using \eq{eq:naiive-defn-energy-change}. For the outcome $(+,e)$, $\Delta E_\rho^\mm(+,e)$ (blue full) is compared to $\Delta \tilde E_\rho^\mm(+,e)$ (red dashed) as $\Delta \theta:= \pi/2 - \theta_2 $ is varied. $\Delta E_\rho^\mm(+,e)= \Delta \tilde E_\rho^\mm(+,e)$ when $\Delta \theta =0$. The inset shows the work statistics  $\mathcal{W}_\rho^\mm(x)$ given a specific measurement model. The work vanishes when the premeasurement unitary commutes with the total Hamiltonian at $\theta_2 = 0$, i.e., $\Delta \theta = \pi/2$.}   \label{fig:Trajectory}
\end{figure}

As an illustrative example for the last case mentioned above, let us consider a system that is a qubit with the time-independent Hamiltonian $H\sub{\s} = (\hbar \omega/2)(|e\>\<e| - |g\>\<g|)$. 
The qubit is initially prepared in the pure state $\ket{\psi} =\ket{\theta_1,+}$, with $\ket{\theta,\pm}:= \pm \cos(\theta/2)\ket{g/e} + \sin(\theta/2)\ket{e/g}$, and is then measured by the sequential observable $\mm := \{\mm_1, \mm_2\}$. The first observable $\mm_1$ is a  projective measurement  with respect to the orthonormal basis  $\ket{\theta_2,\pm}$, while the second observable $\mm_2$  has the POVM elements  $\{M_e^{(2)}, M_g^{(2)}\}$ so that for outcome $e$ ($g$), the qubit is brought to the pure state $\ket{e}$ ($\ket{g}$). The details of this measurement scheme are shown in \app{appendix:Model for a sequence of measurements on a two-level system}. 
In \fig{fig:Trajectory} we plot the conditional change in energy for the sequence of outcomes $x = (+,e)$, $\Delta E_\rho^\mm(+,e)$,  when $\theta_1 = \pi/2$.  
This is generally different to the unconditional energy change $\Delta \tilde E_\rho^\mm(+,e)$, as defined in \eq{eq:naiive-defn-energy-change}, with the two definitions coinciding only when $\theta_2=\theta_1$. 
This is the limit implicitly used in stochastic thermodynamics, where the first observable in the sequence, i.e. $\mm_1$, is ignored. Here, the change in energy is computed given the assumption that  the initial state is $\ket{\theta_1,+}$ and the outcome of the observable $\mm_2$ is $e$. Consequently, the definition for conditional energy change we propose can be seen as a generalisation of the conventional methods used in stochastic thermodynamics, and dispenses with the requirement that the starting point of the trajectory must be a known pure state.
 
\section{Work statistics for measurements}

In order to define the work statistics for a measurement model $\mm$, we must extend the above defined conditional energy increase to the measurement apparatus. The premeasurement and objectification steps of the measurement process can be defined by the  instrument $
\jj_x^\mm(\rho\otimes\xi) :=P\sub{\aa}^x U(\rho\otimes\xi)U^\dagger P\sub{\aa}^x$.
Therefore, by \eq{eq:conditional-energy-change}, the conditional increase in energy of the compound system $\s+\aa$, given outcome $x$ of the POVM $M$, is 
\begin{align}\label{eq:conditional-energy-change-total}
\Delta \e _ \rho^\mm(x) &:= \frac{\tr[(H\sub{\s}(\tau) + H\sub{\aa}(\tau))\jj_x^\mm(\rho\otimes\xi)]}{p_\rho^M(x)}  \nonumber \\
& \qquad - \frac{\mathrm{Re}\left(\tr[\jj_x^\mm((H\sub{\s}(0) + H\sub{\aa}(0))\rho\otimes\xi)]\right)}{p_\rho^M(x)}.
\end{align}

  $\Delta \e _ \rho^\mm(x)$ can be identified as the conditional work, $\w_\rho^\mm(x)$, if it is in principle possible for this quantity to account for all the energy change in the universe as a result of this measurement. We must therefore  be able to  implement an ideal measurement of the observable $Z\sub{\aa} = \sum_{x\in \xx} x P\sub{\aa}^x$ on the apparatus $\aa$, by an inclusion of a second apparatus $\bb$, such that the conditional energy change of the larger system $\s+\aa+\bb$ will  equal that of $\s+\aa$ given by \eq{eq:conditional-energy-change-total}. This way all the energy changes due to measurement will be accounted for, and the source of this energy can be understood as originating from an (implicit) external work storage device, which is outside of the quantum description, the effect of which is the time-dependence of the Hamiltonian of $\s+\aa$.    In \app{appendix:When extending the measurement apparatus will not change the total conditional change in energy} we show that this is possible for all states $\rho$ and observables $M$ if and only if we ensure that $Z\sub{\aa}$ commutes with  the Hamiltonian of $\aa$ when the objectification step of the measurement process is assumed to take place, which is $H\sub{\aa}(\tau)$. This is a consequence of the  Wigner-Araki-Yanase (WAY) theorem \cite{E.Wigner1952,Araki1960,Loveridge2011}.   Given this restriction on $Z\sub{\aa}$,  \eq{eq:conditional-energy-change-total} becomes
\begin{align}\label{eq:conditional-work}
\w_\rho^\mm(x) &= \frac{\tr[P\sub{\aa}^x(H\sub{\s}(\tau) + H\sub{\aa}(\tau)) U(\rho\otimes \xi)U^\dagger] }{p_\rho^M(x)} \nonumber \\
&  - \frac{\mathrm{Re}\left(\tr[ P\sub{\aa}^x U(H\sub{\s}(0) + H\sub{\aa}(0))(\rho\otimes \xi)U^\dagger]\right)}{p_\rho^M(x)}.
\end{align}
The above expression implies that, if $U$ commutes with the total Hamiltonian $H = H\sub{\s}(\tau) + H\sub{\aa}(\tau) = H\sub{\s}(0) + H\sub{\aa}(0) $, then $\w_\rho^\mm(x) = 0$ for all outcomes $x$ and initial states $\rho$. This strengthens previous results, which only established that the average work vanishes when $U$ and $Z\sub{\aa}$ commute with the total Hamiltonian \cite{Mohammady2017}.

In the inset of \fig{fig:Trajectory} we plot $\w_\rho^\mm(x)$ for a specific implementation of the sequential measurement introduced in the previous section (see \app{appendix:Model for a sequence of measurements on a two-level system}). The  work statistics vanishes when $U$ commutes with the total Hamiltonian, achieved at $\theta_2=0$.

\section{Non-recoverable work and the second law of thermodynamics}

In order to relate the obtained work statistics of measurement with the second law of thermodynamics, we make the measurement process cyclic by use of a single thermal bath of  temperature $T$. Namely, at the end of the measurement process we must return the Hamiltonian of system and apparatus to the initial configuration $H\sub{\s}(0) + H\sub{\aa}(0)$, while also returning the state of the compound system to $\rho \otimes \xi$.  The average ``non-recoverable'' or ``irreversible'' work for the measurement process $\mm$ will therefore be given by the average work of performing the measurement, $\<\w_\rho^\mm(x)\> := \sum_{x \in \xx} p_\rho^M(x) \w_\rho^\mm(x)$, plus the \emph{minimal} average work of returning the system to its initial configuration. Below, we shall show that the non-recoverable work will always be non-negative, thus satisfying the second law. 

The average work $\<\w_\rho^\mm(x)\> $ is obtained from 
 \eq{eq:conditional-work} to be $\<\w_\rho^\mm(x)\> = \tr[(U^\dagger H(\tau) U - H(0))(\rho\otimes\xi)]$, where $H(0) = H\sub{\s}(0) + H\sub{\aa}(0)$ and $H(\tau) = H\sub{\s}(\tau) + H\sub{\aa}(\tau)$. Recall that the average state of $\s+\aa$ at the end of the measurement process is $\varrho\sub{\s+\aa}$, defined in \eq{eq:system-apparatus-after-measurement}. Denoting the average state of the system at the end of measurement as  $\rho_\tau:= \tr\sub{\aa}[\varrho\sub{\s+\aa}]$, and the average state of the apparatus at the end of premeasurement and objectification as respectively $\xi':= \tr\sub{\s}[U(\rho\otimes \xi)U^\dagger]$ and $\xi_\tau:= \tr\sub{\s}[\varrho\sub{\s+\aa}]$, the average work can be recast in terms of free energy and information related quantities as 
\begin{align} 
\label{eq:average-work-1}
\<\w_\rho^\mm(x)\> &= \Delta F \sub{\s} +  \Delta F \sub{\aa} +  k_B T\big( I\sub{\s:\aa} +\mathscr{H}   -  \mathscr{X}\sub{\aa}\big), \nonumber \\
& = \Delta F\sub{\s+\aa} + k_B T \big( I\sub{\s:\aa} - I\sub{\s:\aa}' +\mathscr{H}   -  \mathscr{X}\sub{\aa} \big) . 
\end{align}
 Here $k_B$ is Boltzmann's constant; $\Delta F\sub{\s}$ and $\Delta F\sub{\aa}$ are the increase in free energy  of system and apparatus due to the transformations $\rho \mapsto \rho_\tau$ and $\xi \mapsto \xi_\tau$, respectively, while $\Delta F\sub{\s+\aa}$ is the increase in free energy of the compound system due to the transformation $\rho\otimes \xi \mapsto \varrho\sub{\s+\aa} $; $I\sub{\s:\aa}:= S(\rho_\tau) + S(\xi') - S(U(\rho\otimes \xi)U^\dagger)$ is the quantum mutual information between $\s$ and $\aa$ after  premeasurement, where $S(\rho)$ is the von Neumann entropy of $\rho$; $I\sub{\s:\aa}':= S(\rho_\tau) + S(\xi_\tau) - S(\varrho\sub{\s+\aa})$ is the quantum mutual information between $\s$ and $\aa$ after the measurement process;   $\mathscr{H}:= -\sum_{x\in\xx} p_\rho^M(x)\,  \ln{p_\rho^M(x)}$ is the Shannon entropy of the measurement probability distribution $p_\rho^M(x)$; and $\mathscr{X}\sub{\aa}:= S(\xi') - \sum_{x\in \xx} p_\rho^M(x) S(\xi(x))$ is the Holevo information of the apparatus with respect to the ideal measurement of $Z\sub{\aa}$ and its state after premeasurement, $\xi'$. See \app{appendix:Irrecoverable work due to measurement} for details. 

There are two ways of returning the compound system to its initial state $\rho \otimes \xi$ by use of a single thermal bath with temperature $T$: (a) first couple $\s$ with the bath to realise the desired transformation on $\s$, and then repeat the process with $\aa$; or (b) couple the compound system $\s+\aa$ with the bath to bring the total system to the desired state. In case (a) the minimum work cost is  $- \Delta F\sub{\s} - \Delta F\sub{\aa}$, whereas in case (b) the minimum work cost is  $- \Delta F\sub{\s+\aa}$. Both of these are achieved in the limit when the process by which the systems are coupled with the bath is quasi-static \cite{Anders-thermo-discrete}. Therefore, in case (a) the   non-recoverable  work of measurement will be 
\begin{align}\label{eq:non-recoverable-work}
\w_\mathrm{irr}^\mm(\rho) &:= \<\w_\rho^\mm(x)\>- \Delta F\sub{\s} - \Delta F\sub{\aa},\nonumber \\
& =   k_B T\big( I\sub{\s:\aa} +\mathscr{H}   -  \mathscr{X}\sub{\aa} \big),
\end{align}
whereas in case (b) the non-recoverable work will be 
\begin{align}\label{eq:inc-non-recoverable-work}
\w_\mathrm{inc.irr}^\mm(\rho) &:= \<\w_\rho^\mm(x)\> - \Delta F\sub{\s+\aa},\nonumber \\
& =  \w_\mathrm{irr}^\mm(\rho)  -  k_B T \, I\sub{\s:\aa}'. 
\end{align}
Since case (b) uses the correlations between $\s$ and $\aa$ to effect the desired transformation, whereas case (a) does not, we refer to $\w_\mathrm{inc.irr}^\mm(\rho) $ and $\w_\mathrm{irr}^\mm(\rho) $ as the ``inclusive'' and ``non-inclusive'' non-recoverable work, respectively \cite{Manzano2017}.

The non-negativity of $\w_\mathrm{irr}^\mm(\rho)$ follows from the fact that $I\sub{\s:\aa} \geqslant 0$  and $\mathscr{H} \geqslant \mathscr{X}\sub{\aa} \geqslant 0$. Similarly, the non-negativity of $\w_\mathrm{inc.irr}^\mm(\rho)$ can be shown by re-expressing it as $k_B T(S(\varrho\sub{\s+\aa}) - S(\rho\otimes\xi))$, and noting that the  measurement process is unital \cite{DenesPetz2008,Sagawa2012a}. Therefore, the work statistics of measurement as defined by \eq{eq:conditional-work} will  obey the Kelvin statement of the second law \cite{Balian2007}. Moreover, the non-negativity of the mutual information implies that $\w_\mathrm{irr}^\mm(\rho) \geqslant \w_\mathrm{inc.irr}^\mm(\rho)$. We note that the non-inclusive non-recoverable work of measurement is related to the work cost of measurement, $E\sub{\mathrm{cost}}$, discussed in \cite{Sagawa2009b,Abdelkhalek2016a}. Here, only the apparatus is returned to its initial configuration by interacting with the bath. Consequently,  it is related to the non-inclusive non-recoverable work by the equality  $E\sub{\mathrm{cost}} =  \w_\mathrm{irr}^\mm(\rho) + \Delta F\sub{\s}$.

\subsection{Non-recoverable work of projective measurements}
 In \cite{Abdelkhalek2016a} it was shown that the work cost of ideal projective measurements is given as  $E\sub{\mathrm{cost}}^\mathrm{proj} = \tr[H\sub{\s}(\rho_\tau - \rho)] + k_B T \mathscr{H}$, i.e. it only depends on the observable measured and the system states; quantities pertaining to the apparatus do not appear (unlike the case for general observables).  We may generalise this for both the inclusive and non-inclusive non-recoverable work, and for repeatable projective measurements of a sharp observable $M$. Recall that an ideal measurement of $M$ is implemented by the instrument $\ii_x^\mathrm{Ideal}(\rho) = M_x \rho M_x$. A repeatable measurement of $M$ can be implemented by the instrument $\ii_x^\mathrm{Rep}(\rho) = \sum_i q_i V_{x,i}\ii_x^\mathrm{Ideal}(\rho) V_{x,i}^\dagger$, where $q_i$ is a probability distribution, and $V_{x,i}$ are unitaries acting non-trivially only on the subspace projected onto by $M_x$. Such a  measurement is called repeatable because $\tr[\ii_x^\mathrm{Rep}\circ \ii_x^\mathrm{Rep}(\rho)] = \tr[\ii_x^\mathrm{Rep}(\rho)]$, i.e. if we measure the system and obtain outcome $x$, performing the same measurement will reveal $x$ with certainty. Clearly, $\ii_x^\mathrm{Rep} = \ii_x^\mathrm{Ideal}$ if $V_{x,i} = \one$ for all $i$. It follows that the non-recoverable work for repeatable measurements reduces from \eq{eq:non-recoverable-work} and \eq{eq:inc-non-recoverable-work} to
  \begin{align}\label{eq:irr-work-proj}
 \w_\mathrm{irr}^\mathrm{Rep}(\rho) &= k_B T\big(\mathscr{H} + S(\rho_\tau^\mathrm{Rep}) - S(\rho) \big),\nonumber \\
 \w_\mathrm{inc.irr}^\mathrm{Rep}(\rho) &= k_B T \big(S(\rho_\tau^\mathrm{Ideal}) - S(\rho) \big), 
 \end{align}
where $\rho_\tau^\mathrm{Rep} := \sum_{x\in \xx} \ii_x^\mathrm{Rep}(\rho)$ and $\rho_\tau^\mathrm{Ideal} := \sum_{x\in \xx} \ii_x^\mathrm{Ideal}(\rho)$. Note that while the non-inclusive non-recoverable work depends on the particular instrument $\ii_x^\mathrm{Rep}$, the same is not true for the inclusive non-recoverable work. The reason that \eq{eq:irr-work-proj} is obtained is that for repeatable measurements, $S(\xi(x)) = S(\xi)$ for all $x \in \xx$ (see \app{appendix:Non-recoverable work due to projective  measurements}).

As discussed in \cite{Allahverdyan2011a,Guryanova2018},  ideal (or repeatable) projective measurements require that the initial apparatus state $\xi$ not have full rank.  Therefore, by the third law of thermodynamics, such  measurements will require infinite resources \cite{Allahverdyan2011a, Wu2013, Reeb2013a, Masanes2014}. To account for this, \cite{Guryanova2018} introduced  measurement models of a PVM wherein the apparatus is initially in a thermal state of finite temperature. These measurements will necessarily be unrepeatable. We show in \app{appendix:Non-recoverable work due to projective  measurements} that the non-recoverable work for projective measurements that use a full-rank apparatus of dimension $\dim(\h\sub{\aa}) = \dim(\h\sub{\s})$, which we call noisy projective measurements, is of the form
\begin{align}\label{eq:work-noisy} 
\w_\mathrm{irr}^\mathrm{Noisy}(\rho) &= \w_\mathrm{inc.irr}^\mathrm{Noisy}(\rho)= k_B T \big(S(\rho_\tau^\mathrm{Ideal}) - S(\rho) \big).
\end{align}  
Here, the premeasurement unitary is a SWAP operation, followed by an appropriate local unitary on $\aa$. Consequently the mutual information terms in \eq{eq:non-recoverable-work} and \eq{eq:inc-non-recoverable-work} vanish, and we are left with $\mathscr{H} - \mathscr{X}\sub{\aa}  = S(\rho_\tau^\mathrm{Ideal}) - S(\rho)$. As with the case of ideal and repeatable measurements, the non-recoverable work for such noisy measurements is purely determined by the instrument implementing $M$, and $\rho$; the non-recoverable work is independent of the apparatus state. However, since noisy measurements require the apparatus to have the same Hilbert space as the system, the independence of the non-recoverable work on the apparatus is not as general as the case for repeatable measurements.

Finally, comparing all three classes of projective measurements, we observe the following  relations:
\begin{align}
\w_\mathrm{irr}^\mathrm{Rep}(\rho)& \geqslant \w_\mathrm{irr}^\mathrm{Ideal}(\rho) \geqslant \w_\mathrm{irr}^\mathrm{Noisy}(\rho), \nonumber \\
\w_\mathrm{inc.irr}^\mathrm{Rep}(\rho)& = \w_\mathrm{inc.irr}^\mathrm{Ideal}(\rho) = \w_\mathrm{inc.irr}^\mathrm{Noisy}(\rho).
\end{align} 
The first inequality in the top line is due to the random unitary mixing in $\ii_x^\mathrm{Rep}$, with the equality condition being satisfied if $V_{x,i} = V_x$ for all $i$ and $x$. Meanwhile, the second inequality in the top line follows from the fact that $ \w_\mathrm{irr}^\mathrm{Ideal}(\rho) - \w_\mathrm{irr}^\mathrm{Noisy}(\rho) =  k_B T\, \mathscr{H}$, and so the equality condition is only obtained if $\mathscr{H}=0$. However, these inequalities are replaced with  equalities on the bottom line, showing that the inclusive non-recoverable work of repeatable, ideal,  and noisy measurements is identical.

\section{Conclusions}

In the present work we have defined the change of energy, conditional on the outcome of a general quantum measurement, as the difference in conditional energies of the system, evaluated before and after the measurement process. In order for this definition to be physically meaningful, we imposed three requirements on the definitions for the initial and final conditional energies.   We showed that the most general expression for the conditional energy of the system, after the measurement process, is simply the expected value of the Hamiltonian, given the transformed state of the system after a  measurement outcome. This is as one would expect. However the most general, and operationally meaningful, expression for the conditional energy of the system, evaluated before the measurement process, is not the expected value of the Hamiltonian given the initial state, but rather the real component of the weak value of the Hamiltonian.   We have shown that our definition  provides a unified platform which, in specific cases, reproduces previously existing results such as the TPM or   quasi-probability distributions for work in isolated systems, and energy change along pure-state quantum trajectories. By extending our definition to the apparatus used to measure an observable, we  determine the work statistics for general quantum measurements, and show that these vanish for all measurement outcomes when the measurement process conserves energy. Finally, to link the work distribution to the second law of thermodynamics, we characterise the non-recoverable work when the measurement process is embedded within a cyclic heat engine in contact with a single thermal bath. The non-recoverable work is non-negative, thus satisfying the second law. Moreover, we show that the non-recoverable work is purely determined by system quantities for two classes of projective measurements: repeatable projective measurements, and non-repeatable, noisy projective measurements. Here, noisy projective measurements are implemented by use of an apparatus that has the same dimension as the system, and is initially in a state of full-rank.    

\acknowledgements

The authors thank K. V. Hovhannisyan, H. J. D. Miller,  C. Elouard and L. Loveridge for their useful comments. We acknowledge research support from EPSRC (Grant EP/P030815/1).


\bibliographystyle{apsrev4-1}
\bibliography{Thermodynamics-Thermal-Measurements}

\newpage

\appendix

\onecolumngrid


\makeatletter
\renewcommand\p@subsection{\thesection\,}
\makeatother
\makeatletter
\renewcommand\p@subsubsection{\thesection\,\thesubsection\,}
\makeatother

%

\section{General form of conditional energy change}
\label{appendix:proof-of-weak-value}

We show here that   \eq{eq:general-form-f} and  \eq{eq:energy-after} are the most general forms of the average conditional energies $E_{x,0}^\mm(\rho)$ and $E_{x,\tau}^\mm(\rho)$, respectively, that fulfill all three physical requirements.  To this aim, first it is easy to show that a function  $E_{x,t}^\mm(\rho)$ will satisfy both \reqref{requirement1} and \reqref{requirement2} if and only if it can be written as
\begin{align}\label{eq:general-form-E-f}
E_{x,t}^\mm(\rho) = \frac{f_{x,t}^\mm(\rho)}{p_\rho^M(x)},
\end{align}
such that (i) $f_{x,t}^\mm(\rho)$ is linear in $\rho$, and (ii)  $\sum_{x\in \xx} f_{x,0}^\mm(\rho) = \tr[H\sub{\s}(0) \rho]$, while $\sum_{x\in \xx} f_{x,\tau}^\mm(\rho) = \sum_{x\in \xx} \tr[H\sub{\s}(\tau) \ii_x^\mm(\rho)] =: \tr[H\sub{\s}(\tau) \rho_\tau]$.

The if statement of the proof is trivial. To prove the only if statement,  we first note that \eq{eq:cond-E-convex} can be re-expressed as $\sum_k p_k (p_{\rho^{(k)}}^M(x) E_{x,t}^\mm(\rho^{(k)}))  = p_\rho^M(x) E_{x,t}^\mm(\rho) $. By introducing the relabeling   $f_{x,t}^\mm(\rho) := p_{\rho}^M(x) E_{x,t}^\mm(\rho) $,   \eq{eq:cond-E-convex} can therefore be rewritten as $\sum_k p_k f_{x,t}^\mm(\rho^{(k)}) = f_{x,t}^\mm(\rho)$. It follows that in order for \reqref{requirement2} to be satisfied, we must have (i). Finally, inserting $f_{x,t}^\mm(\rho) := p_{\rho}^M(x) E_{x,t}^\mm(\rho)$ in \eq{eq:req1-avg-cond}  proves that in order for \reqref{requirement1} to also be satisfied, we must have (ii).  

Now we turn to \reqref{requirement3}. The most general form of $E_{x,t}(\rho)$ that can be written in the form of \eq{eq:general-form-E-f} while also satisfying  \reqref{requirement3}, can be expressed as
\begin{align}\label{eq:general-req-3}
E_{x,t}(\rho) = \frac{\gamma \,  \tr[H\sub{\s}(t) \ii_x^\mm(\rho) ] + (1-\gamma)\, \left(\lambda \, \tr[\ii_x^\mm(H\sub{\s}(t) \rho) ] +  (1-\lambda) \, \tr[\ii_x^\mm(H\sub{\s}(t) \rho) ]^* \right)}{p_\rho^M(x)}, 
\end{align}
where $\gamma, \lambda \in [0,1]$.  Clearly, the numerator in \eq{eq:general-req-3} is linear in $\rho$ thus satisfying (i), and when $\tr[\ii_x^\mm(P\sub{\s}^j(t)\rho P\sub{\s}^j(t))] = \tr[P\sub{\s}^j(t) \ii_x^\mm(\rho)]$, \eq{eq:general-req-3} will reduce to $\tr[H\sub{\s}(t) \rho(x)]$, thus satisfying \reqref{requirement3}. To see this, we note that given this relation, it follows that 
\begin{align}
 \tr[\ii_x^\mm(H\sub{\s}(t) \rho) ] &= \sum_j \epsilon_j(t)\tr[\ii_x^\mm(P\sub{\s}^j(t)\rho)], \nonumber \\
 & = \sum_{j,k} \epsilon_j(t)\tr[P\sub{\s}^k(t)\ii_x^\mm(P\sub{\s}^j(t)\rho)], \nonumber \\
 & = \sum_{j} \epsilon_j(t)\tr[\ii_x^\mm(P\sub{\s}^j(t)\rho P\sub{\s}^j(t))], \nonumber \\
 &= \sum_j \epsilon_j(t)\tr[P\sub{\s}^j(t)\ii_x^\mm(\rho)] =  \tr[H\sub{\s}(t) \ii_x^\mm( \rho) ].
\end{align}
Finally, in order for \eq{eq:general-req-3} to satisfy (ii), thus fulfilling all three physical requirements, we must have $\gamma = 0$ for $t=0$, and $\gamma = 1$ for $t = \tau$. This is because $\sum_{x\in \xx} \tr[H\sub{\s}(t) \ii_x^\mm(\rho) ] = \tr[H\sub{\s}(t) \rho_\tau]$, whereas 
\begin{align}
\sum_{x\in \xx} \lambda \, \tr[\ii_x^\mm(H\sub{\s}(t) \rho) ] +  (1-\lambda) \, \tr[\ii_x^\mm(H\sub{\s}(t) \rho) ]^* = \tr[H\sub{\s}(t) \rho].
\end{align} 
Therefore, the most general expression for the conditional energies which satisfy all three physical requirements are
\begin{align}
E_{x,0}^\mm(\rho) &= \frac{\lambda \, \tr[\ii_x^\mm(H\sub{\s}(0) \rho) ] +  (1-\lambda) \, \tr[\ii_x^\mm(H\sub{\s}(0) \rho) ]^*}{p_\rho^M(x)}, \nonumber \\
&\equiv  \lambda \, {}_x\langle H\sub{\s}(0) \rangle_\rho + (1-\lambda) \, {}_x\langle H\sub{\s}(0) \rangle_\rho^*,
\end{align}
and
\begin{align}
E_{x,\tau}^\mm(\rho) = \tr[H\sub{\s}(\tau) \rho(x) ].
\end{align}

\section{Conditional change in energy given multiple measurements}\label{Conditional change in energy given multiple measurements}

In many  situations, a system may be subject to a sequence of measurements. However, care must be taken when considering the corresponding conditional energy changes, since these are not additive when only one measurement is considered at a time. Specifically, consider the measurement  $\mm:= \{\mm_i\}_{i=1}^I$, with the corresponding outcomes $x:= \{x_i\}_{i=1}^I$. Here, the system is sequentially measured by the observables $\mm_i$, with the recorded outcomes $x_i \in \xx_i$. Given the initial preparation of the system as $\rho$, and assuming that the Hamiltonians before and after  the $i$\ts{th} measurement are $H\sub{\s}(t_i)$ and $H\sub{\s}(t_{i+1})$ respectively, the corresponding energy change given the full sequence of measurements is given as $\Delta E_\rho^{\mm}(x) = E_{x,t_{I+1}}^{\mm}(\rho)- E_{x,t_1}^{\mm}(\rho) = \tr[H\sub{\s}(t_{I+1}) \rho(x_I)] - \mathrm{Re}(\tr[\ii_x^\mm(H\sub{\s}(t_1) \rho)])/\tr[\ii_x^\mm(\rho)]$, where   $\rho(x_i)$ is the normalised state of $\s$ after the $i$\ts{th} measurement in the sequence, and $\ii_x^\mm = \ii_{x_{I}}^{\mm_{I}}\circ \dots \circ \ii_{x_{1}}^{\mm_{1}}$ the sequential application of instruments $\ii_{x_{i}}^{\mm_{i}}$.  Denoting  $\rho \equiv \rho(x_0)$, the sum of conditional energy changes for each measurement  will be 
\begin{align}
\sum_{i=1}^I \Delta E_{\rho(x_{i-1})}^{\mm_i}(x_i) &= \sum_{i=1}^I \tr[H\sub{\s}(t_{i+1}) \rho(x_i)] \nonumber \\ 
& - \sum_{i=1}^I \mathrm{Re}\left(\frac{\tr[\ii_{x_i}^{\mm_i}(H\sub{\s}(t_{i}) \rho(x_{i-1}))]}{\tr[\ii_{x_i}^{\mm_i}(\rho(x_{i-1})]}\right), \nonumber \\
& \ne \Delta E_\rho^{\mm}(x).
\end{align}
The reason for the non-additivity is two-fold. First,  the final energy after measuring $\mm_i$, namely, $\tr[H\sub{\s}(t_{i+1}) \rho(x_i)]$,  does not equal the initial energy before measuring $\mm_{i+1}$, namely, $\mathrm{Re}(\tr[\ii_{x_{i+1}}^{\mm_{i+1}}(H\sub{\s}(t_{i+1}) \rho(x_{i}))])/ \tr[\ii_{x_{i+1}}^{\mm_{i+1}}(\rho(x_{i}))]$. Secondly, the weak value of the Hamiltonian evaluated at any point in the sequence of measurements depends on the full sequence of measurements performed afterwards, and not just the measurement performed immediately after.  

In order to have an additive notion of conditional energy changes, we must extend our definition as follows: first,  augment the sequence of measurements by including a trivial measurement at the end, i.e., let $\mm:= \{\mm_i\}_{i=1}^{I+1}$, such that $\mm_{I+1}$ is a trivial observable that does not reveal any information about the system, and is equivalent to not performing any measurement at all. In other words let $\mm_{I+1}$ have only one outcome $x_{I+1}$, with the instrument satisfying   $\tr[\ii_{x_{I+1}}^{\mm_{I+1}}(A)] = \tr[A]$. The conditional energy of the system, before measuring the $i$\ts{th} observable in the sequence, can thus be given as 
\begin{align}\label{eq:conditional-energy-sequence}
E_{x_{i}, t_{i}}^{\mm_{i}}(\rho) := \frac{\mathrm{Re}(\tr[(\ii_{x_{I+1}}^{\mm_{I+1}}\circ \dots \circ \ii_{x_{i}}^{\mm_{i}} ) (H\sub{\s}(t_{i}) \rho(x_{i-1})) ])}{\tr[(\ii_{x_{I+1}}^{\mm_{I+1}}\circ \dots \circ \ii_{x_{i}}^{\mm_{i}} )  \rho(x_{i-1}))]}.
\end{align}  
 Note that  $E_{x_{I+1}, t_{I+1}}^{\mm_{I+1}}(\rho) = E_{x,t_{I+1}}^\mm$, and $E_{x_{1}, t_{1}}^{\mm_{1}}(\rho) = E_{x,t_1}^\mm(\rho)$.
By defining the change in conditional energy during the $i$\ts{th} measurement as $\Delta E_\rho^{\mm_i}(x_i) := E_{x_{i+1}, t_{i+1}}^{\mm_{i+1}}(\rho) - E_{x_{i}, t_{i}}^{\mm_{i}}(\rho)$, we therefore obtain an additive notion of conditional energy change which satisfies $\sum_{i=1}^{I} \Delta E_\rho^{\mm_i}(x_i) = \Delta E_\rho^{\mm}(x)$. 

For long sequences of measurements, such as those encountered in stochastic quantum thermodynamics, calculating \eq{eq:conditional-energy-sequence} can be  very arduous. However, as shown in \app{appendix:Connection to quantum stochastic thermodynamics}, if the measurement of $\mm_i$ first projects $\rho(x_{i-1})$ onto one of its pure state components $\ket{\psi(x_{i-1})}$, then  \eq{eq:conditional-energy-sequence} is reduced to the simple expression $\<\psi(x_{i-1})|H\sub{\s} |\psi(x_{i-1}) \>$.

\section{Limiting cases of conditional energy change}\label{appendix:Limiting case of conditional work statistics}
\subsection{Connection to the TPM protocol}\label{appendix:Connection to the TPM protocol}
Consider the measurement model $\mm$ which induces the instruments $\ii_{(m,n)}^\mm(\rho) = P\sub{\s}^n(\tau) U P\sub{\s}^m(0) \rho P\sub{\s}^m(0) U^\dagger P\sub{\s}^n(\tau)$, and the associated effect operators $M_{(m,n)}:= P\sub{\s}^m(0) U^\dagger P\sub{\s}^n(\tau) U P\sub{\s}^m(0)$ for the POVM $M$ with outcomes $x:=(m,n)$. Here, $U$ is a unitary operator, while $P\sub{\s}^n(t)$ are time-local spectral projections of a Hamiltonian $H\sub{\s}(t)$, with energy eigenvalues $\epsilon_n(t)$. By \eq{eq:conditional-energy-change} and \eq{eq:cond-E-weak}, the conditional change in energy for each outcome  can thus be written as 
\begin{align}\label{eq:TPM-work-dist}
\Delta E_\rho^\mm(m,n) &=\frac{\tr[H\sub{\s}(\tau) \ii_{(m,n)}^\mm(\rho)]}{p_\rho^M(m,n)} - \frac{\mathrm{Re}\left(\tr[\ii_{(m,n)}^\mm(H\sub{\s}(0) \rho)]\right)}{p_\rho^M(m,n)} , \nonumber \\
& = \epsilon_n(\tau) - \epsilon_m(0),
\end{align}
which is a difference in energy eigenvalues. This is  the work distribution one obtains from the TPM protocol, wherein a  quantum system undergoing unitary evolution $U$ generated by its Hamiltonian is projectively measured by its time-local Hamiltonian at two points of its evolution.  The average change in energy, over all measurement outcomes, is thus given by the probabilities $p_\rho^M(m,n) := \tr[P\sub{\s}^m(0) U^\dagger P\sub{\s}^n(\tau) U P\sub{\s}^m(0) \rho]$ to be
\begin{align}\label{eq:avg-work-TPM}
\sum_{m,n}p_\rho^M(m,n) \Delta E_\rho^\mm(m,n) = \sum_m \tr[H\sub{\s}(\tau) U P\sub{\s}^m(0) \rho  P\sub{\s}^m(0) U^\dagger] - \tr[H\sub{\s}(0) \rho].
\end{align}

\subsection{Connection to quasi-probability distributions of work in isolated systems}\label{appendix:Connection to quasi-probability distributions of work}

As is well known, \eq{eq:avg-work-TPM} will fail to equal the increase in average energy of an isolated system as it unitarily evolves by $U$, namely $\tr[H\sub{\s}(\tau) U \rho U^\dagger] - \tr[H\sub{\s}(0) \rho]$, except for when $\rho$ commutes with $H\sub{\s}(0)$. Let us therefore consider a different measurement model $\mm$, which induces the instruments $\ii_{n}^\mm(\rho) = P\sub{\s}^n(\tau) U  \rho  U^\dagger P\sub{\s}^n(\tau)$ with the associated effect operators  $M_n := U^\dagger P\sub{\s}^n(\tau) U $ for the POVM $M$ with outcomes $n$. This is equivalent to the ideal measurement of the Heisenberg evolved Hamiltonian $\tilde H\sub{\s}(\tau):= U^\dagger H\sub{\s}(\tau) U$, with spectral projections $\tilde P\sub{\s}^n(\tau) := U^\dagger P\sub{\s}^n(\tau) U$. The conditional change in energy now becomes
\begin{align}\label{eq:OPM-work-dist}
\Delta E_\rho^\mm(n) &=\frac{\tr[H\sub{\s}(\tau) \ii_n^\mm(\rho)]}{p_\rho^M(n)} - \frac{\mathrm{Re}\left(\tr[\ii_n^\mm(H\sub{\s}(0) \rho)] \right)}{p_\rho^M(n)} , \nonumber \\
& = \epsilon_n(\tau) - \frac{\mathrm{Re}\left(\tr[\tilde P\sub{\s}^n(\tau) H\sub{\s}(0) \rho] \right)}{p_\rho^M(n)}, 
\end{align}
which always averages to 
\begin{align}\label{eq:avg-work-OPM}
\sum_{n}p_\rho^M(n) \Delta E_\rho^\mm(n) &=  \tr[H\sub{\s}(\tau) U \rho U^\dagger] - \tr[H\sub{\s}(0) \rho].
\end{align}
 However, this comes at the expense of the energy change to no longer necessarily be equal to the difference in energy eigenvalues when   $\rho$ commutes with $H\sub{\s}(0)$.  This is because unless $\rho$ only has support on a single energy subspace, then the conditional initial energy in \eq{eq:OPM-work-dist} will be an average over energy eigenvalues.  Indeed, in order for $\mathrm{Re}\left(\tr[M_x H\sub{\s}(0) \rho]\right)/p_\rho^M(x) = \epsilon_m(0)$ to obtain for any state $\rho$ that commutes with $H\sub{\s}(0)$, then it is necessary for the support of $M_x$ to be contained in the  energy subspace of $\epsilon_m(0)$. It will follow that  $M_x$ would be equivalent to $P\sub{\s}^m(0) M_x P\sub{\s}^m(0)$, resulting in the instruments $\ii_x^\mm(\rho) = \ii_x^\mm(P\sub{\s}^m(0) \rho P\sub{\s}^m(0))$ and, hence, the channel  $\ii^\mm(\rho) : = \sum_{x\in \xx}\ii_x^\mm(\rho) = \sum_m \ii^\mm(P\sub{\s}^m(0) \rho P\sub{\s}^m(0))$. We may conclude that the channel $\ii^\mm$ cannot model the unitary evolution of a system with a state $\rho$ that does not commute with $H\sub{\s}(0)$. As such, when the state $\rho$ does not commute with the Hamiltonian, it follows that the measurement  will not result in an average work that equals $\tr[H\sub{\s}(\tau) U \rho U^\dagger] - \tr[H\sub{\s}(0) \rho]$. In other words a measurement model $\mm$ that obtains the TPM statistics of work, when the system is prepared as a mixture of energy eigenstates,  cannot give the average work as being the increase in average energy, as the system transforms unitarily, for arbitrary states.  Our definition for conditional energy change is therefore in full agreement with the no-go theorem of \cite{Perarnau-Llobet2016a}. 

 We now show that the generalisation of conditional energy change beyond energy eigenvalue differences is equivalent to the generalisation of probabilities to quasiprobabilities introduced in \cite{Allahverdyan2014}. The summand  in \eq{eq:avg-work-OPM} can be written as  
\begin{align}
p_\rho^M(n) \Delta E_\rho^\mm(n) &= \epsilon_n(\tau) \tr[ \tilde  P\sub{\s}^n(\tau)  \rho ] - \sum_m \epsilon_m(0) \mathrm{Re}\left(\tr[ \tilde P\sub{\s}^n(\tau)  P\sub{\s}^m(0) \rho]\right), \nonumber \\
& =  \sum_m \epsilon_n(\tau) \mathrm{Re}\left(\tr[ \tilde  P\sub{\s}^n(\tau) P\sub{\s}^m(0) \rho ]\right) \nonumber \\
&\qquad -\sum_m  \epsilon_m(0) \mathrm{Re}\left(\tr[ \tilde P\sub{\s}^n(\tau)  P\sub{\s}^m(0) \rho]\right) , \nonumber \\
& = \sum_m \tilde p_{(m,n)}(\epsilon_n(\tau) - \epsilon_m(0) ).
\end{align}
On the left hand side, we have a real probability distribution $p_\rho^M(n)$, and  energy change $\Delta E_\rho^\mm(n)$ which may lie outside the range of energy eigenvalue differences. Meanwhile, on the right hand side, the energy change of $\s$ is  defined with respect to the random variable $\epsilon_n(\tau) - \epsilon_m(0)$, which is sampled by the quasi-probability distribution $\tilde p_{(m,n)}:= \mathrm{Re}\left(\tr[ \tilde P\sub{\s}^n(\tau)  P\sub{\s}^m(0) \rho]\right)$.

\subsection{Connection to quantum stochastic thermodynamics}\label{appendix:Connection to quantum stochastic thermodynamics}

Consider a quantum state  $\rho = \sum_m q_m \prs{\psi_m}$, where $\prs{\psi} \equiv |\psi\>\<\psi|$ is a projection on the vector $\ket{\psi} \in \h\sub{\s}$. We may define a quantum trajectory induced by the POVM $M := \{M^{(1)}, M^{(2)}\}$ as the sequence of measurement outcomes $x:= (m,n)$, with $m$ the outcome of an ideal projective measurement  of the observable $M^{(1)} :=\sum_m m\prs{\psi_m}$,  and $n$ the outcome of a subsequent measurement of a general POVM $M^{(2)}$. Given that the effect operators for the POVM $M$ defined by the sequence of measurements are $M_x := \prs{\psi_m} M_n^{(2)} \prs{\psi_m}$,   the conditional initial energy of the system is given by \eq{eq:cond-E-weak} as
\begin{align}
E_{x,0}^\mm(\rho) &= \frac{\mathrm{Re}\left(\tr[ \prs{\psi_m}M_n^{(2)} \prs{\psi_m} H\sub{\s}(0) \rho]\right)}{p_\rho^M(x)}, \nonumber \\
&= \frac{q_m \<\psi_m|M_n^{(2)}|\psi_m\>\<\psi_m|H\sub{\s}(0)|\psi_m\>}{q_m \<\psi_m|M_n^{(2)}|\psi_m\>}, \nonumber \\
& = \<\psi_m|H\sub{\s}(0)|\psi_m\>.
\end{align}
Therefore, the conditional initial energy of the system, along the quantum trajectory $x=(m,n)$, is the expected energy of the system when it is in the pure state $\ket{\psi_m}$ that defines the starting point of the trajectory in question. We note that this result holds even when $\rho$ is already a pure state $\ket{\psi_m}$.  Of course, the preparation of a quantum system in a pure state can also be recast in terms of projective measurements. Cast in this light, this means that in order for $E_{x,0}^\mm(\rho) = \<\psi_m| H\sub{\s}(0)|\psi_m\>$ to hold, then $M$ must also account for the preparation of the system in state $\ket{\psi}$. This should not be too surprising, however, since when we say that we ``know'' that the system starts in a given pure state $\ket{\psi_m}$, operationally this means that we must have performed a projective measurement to verify this fact.

\section{When extending the measurement apparatus will not change the total conditional increase in energy}\label{appendix:When extending the measurement apparatus will not change the total conditional change in energy}

Let us augment the measurement model  $\mm:= (\h\sub{\aa}, \xi\sub{\aa}, U, Z\sub{\aa})$ by including a measurement model for $Z\sub{\aa}$. The augmented model is denoted   $\mm':= (\h\sub{\aa}\otimes  \h\sub{\bb} , \xi\sub{\aa}\otimes \xi\sub{\bb}, VU , Z\sub{\bb})$. Here $\xi\sub{\bb}$ is the initial state of apparatus $\bb$, and  $V$ is a unitary operator that acts on $\h\sub{\aa}\otimes \h\sub{\bb}$ after the application of $U$ on $\h\sub{\s}\otimes \h\sub{\aa}$. Finally, $Z\sub{\bb}=\sum_{x\in \xx} x P\sub{\bb}^x$ is the observable on apparatus $\bb$. Denoting $H(t):= H\sub{\s}(t)+H\sub{\aa}(t) + H\sub{\bb}(t)$ for $t\in \{0, \tau\}$, and remembering that $\mm'$  and $\mm$ are models for the same POVM $M$ and, hence, give the same probabilities, we obtain the conditional energy increase of the compound system  by generalising \eq{eq:conditional-energy-change-total} as

\begin{align}\label{eq:conditional-energy-change-augmented}
\Delta \e _ \rho^{\mm'}(x) &:= \frac{\tr[H(\tau) P\sub{\bb}^xVU(\rho\otimes\xi\sub{\aa}\otimes \xi\sub{\bb})U^\dagger V^\dagger P\sub{\bb}^x]}{p_\rho^M(x)}  \nonumber \\
& - \frac{\mathrm{Re}\left(\tr[ P\sub{\bb}^xVU H(0)( \rho\otimes\xi\sub{\aa}\otimes \xi\sub{\bb})U^\dagger V^\dagger ]\right)}{p_\rho^M(x)}.
\end{align}
We now wish to show that it is possible to have $\Delta \e _ \rho^{\mm'}(x) = \Delta \e _ \rho^{\mm}(x)$ for all states $\rho$ if $Z\sub{\aa}$ commutes with $H\sub{\s}(\tau)$. Let us assume that this commutation relation holds.   By the WAY theorem \cite{E.Wigner1952,Araki1960,Loveridge2011}, it follows that we may perform an ideal measurement of $Z\sub{\aa}$  by a unitary coupling $V$  with an apparatus $\bb$, followed by an ideal measurement of this apparatus by $Z\sub{\bb}$,  such that both  $V$ and  $Z\sub{\bb}$ commute with $H\sub{\aa}(\tau) + H\sub{\bb}(\tau)$. If these conditions, together with $H\sub{\bb}(\tau) = H\sub{\bb}(0)$, are satisfied, then \eq{eq:conditional-energy-change-augmented} can be shown to reduce to 

\begin{align}
\Delta \e _ \rho^{\mm'}(x) &=  \frac{\tr[V^\dagger P\sub{\bb}^xV(H\sub{\s}(\tau) + H\sub{\aa}(\tau))U(\rho\otimes\xi\sub{\aa} )U^\dagger \otimes \xi\sub{\bb} ]}{p_\rho^M(x)}  \nonumber \\
& \, \, - \frac{\mathrm{Re}\left(\tr[ V^\dagger P\sub{\bb}^xVU(H\sub{\s}(0) + H\sub{\aa}(0))(\rho\otimes\xi\sub{\aa} )U^\dagger \otimes \xi\sub{\bb} ]\right)}{p_\rho^M(x)} , \nonumber \\
& +  \frac{\tr[V^\dagger P\sub{\bb}^xVU(\rho\otimes\xi\sub{\aa} )U^\dagger \otimes (H\sub{\bb}(\tau) - H\sub{\bb}( 0))\xi\sub{\bb} ]}{p_\rho^M(x)} \nonumber \\ 
& = \frac{\tr[P\sub{\aa}^x(H\sub{\s}(\tau) + H\sub{\aa}(\tau))U(\rho\otimes\xi\sub{\aa})U^\dagger ]}{p_\rho^M(x)}  \nonumber \\
& \, \,  - \frac{\mathrm{Re}\left(\tr[ P\sub{\aa}^x U(H\sub{\s}(0) + H\sub{\aa}(0))(\rho\otimes\xi\sub{\aa})U^\dagger  ]\right)}{p_\rho^M(x)} = \Delta \e_\rho^\mm(x)
\end{align}
for all $\rho$. In the last line, we have used the fact that  $\tr[V^\dagger P\sub{\bb}^xV (O\sub{\s+\aa}\otimes \xi\sub{\bb})] = \tr[P\sub{\aa}^x O\sub{\s+\aa}]$ by construction. Therefore, if $Z\sub{\aa}$ commutes with $H\sub{\aa}(\tau)$, then it is possible to extend the measurement model while not changing the total conditional increase in energy  due to measurement. 

To show that $Z\sub{\aa}$ commuting with $H\sub{\aa}(\tau)$ is the only possibility for $\Delta \e _ \rho^{\mm'}(x) = \Delta \e_\rho^\mm(x)$ to hold for all $\rho$, let us look at a counter example. By the WAY theorem, if $Z\sub{\aa}$ does not commute with $H\sub{\aa}(\tau)$, then an ideal measurement of this observable precludes the possibility that $V$ commutes with $H(\tau)$. As such, even if we choose $Z\sub{\bb}$ to commute with $H(\tau)$, it will generally not be possible to reduce \eq{eq:conditional-energy-change-augmented} to \eq{eq:conditional-energy-change-total}. In such cases, the energy change of $\s+\aa$ cannot account for the total energy change due to measurement.

\section{Model for a sequence of measurements on a two-level system}\label{appendix:Model for a sequence of measurements on a two-level system}

Let us consider a two-level system $\s$  with Hamiltonian $H\sub{\s} = \frac{\hbar \omega}{2}(\prs{e} - \prs{g})$. Let the system interact sequentially with two apparatus systems $\aa$ and $\bb$, with Hilbert spaces $\h\sub{\aa} \simeq \co^2$ and $\h\sub{\bb} \simeq \co^4$, and Hamiltonians $H\sub{\aa} = \one\sub{\aa}$ and $H\sub{\bb} = \hbar \omega \sum_{n=0}^3 P\sub{\bb}[n]$, respectively. Defining the angle-dependent  states $\ket{\theta,\pm}:= \pm \cos(\theta/2) \ket{g/e} + \sin(\theta/2) \ket{e/g}$ for the two-level system,  let us define the two premeasurement unitary operators
\begin{align}
U_{\s+\aa}:= \ket{\theta_2,\pm}\otimes \ket{\phi} \mapsto \ket{\theta_2,\pm}\otimes \ket{\phi_\pm},
\end{align}
and
\begin{align}
U_{\s+\bb}:=\begin{cases}
    \ket{g}\otimes \ket{1} \mapsto \ket{e}\otimes \ket{0},  \\
    \ket{e}\otimes \ket{1} \mapsto \ket{e}\otimes \ket{1},  \\
    \ket{g}\otimes \ket{2} \mapsto \ket{g}\otimes \ket{2},  \\
    \ket{e}\otimes \ket{2} \mapsto \ket{g}\otimes \ket{3}.  
  \end{cases}
\end{align}
The full premeasurement unitary is then given as $U = U_{\s+\bb}U_{\s+\aa}$. Let the systems
$\aa$ and $\bb$ be initially prepared in states $\xi\sub{\aa} = \pra{\phi}$ and $\xi\sub{\bb} = q P\sub{\bb}[1] + (1-q) P\sub{\bb}[2]$, respectively. Finally, let the apparatus observables be $Z\sub{\aa} = \pra{\phi_+} - \pra{\phi_-}$ and $Z\sub{\bb} = e P\sub{\bb}^e + gP\sub{\bb}^g$, where  $ P\sub{\bb}^e = P\sub{\bb}[0] + P\sub{\bb}[1] $ and $ P\sub{\bb}^g = P\sub{\bb}[2] + P\sub{\bb}[3]$. 

This model defines the POVM $M_{m,n}$, where $M_{\pm,e}:= q\prs{\theta_2, \pm}$ and $M_{\pm,g}:= (1-q)\prs{\theta_2, \pm}$, with the post-measurement states of $\s$ being given as $\rho(\pm,e) = \prs{e}$ and $\rho(\pm,g) = \prs{g}$. 

By \eq{eq:conditional-energy-change}, the conditional change in energy of $\s$ is  given as 
\begin{align}
\Delta E_\rho^\mm(\pm, e/g) &:= \frac{\tr[H\sub{\s} \pra{\pm}P\sub{\bb}^{e/g}U(\rho\otimes\xi\sub{\aa}\otimes \xi\sub{\bb})U^\dagger]}{p_\rho^M(\pm,e/g)} \nonumber \\
&-\frac{\mathrm{Re}\left(\tr[ \pra{\pm}P\sub{\bb}^{e/g}U(H\sub{\s}\rho\otimes\xi\sub{\aa}\otimes \xi\sub{\bb})U^\dagger]\right)}{p_\rho^M(\pm,e/g)},
\end{align}
and similarly the work is given by \eq{eq:conditional-work} as
\begin{align}
\w_\rho^\mm(\pm, e/g) &:= \frac{\tr[H \pra{\pm}P\sub{\bb}^{e/g}U(\rho\otimes\xi\sub{\aa}\otimes \xi\sub{\bb})U^\dagger]}{p_\rho^M(\pm,e/g)} \nonumber \\
&-\frac{\mathrm{Re}\left(\tr[ \pra{\pm}P\sub{\bb}^{e/g}U(H\rho\otimes\xi\sub{\aa}\otimes \xi\sub{\bb})U^\dagger]\right)}{p_\rho^M(\pm,e/g)},
\end{align}
where $H = H\sub{\s} + H\sub{\aa} + H\sub{\bb}$ is the total Hamiltonian. It is simple to verify that when $\theta_2 = 0$, then $U$ commutes with $H$, and so the work values for all four outcomes will vanish. \fig{fig:Trajectory} shows the conditional energy change and work statistics for this measurement model, when the initial system state is chosen as $\rho = \prs{\theta_1,+}$, with  $\theta_1 = \pi/2$.

\section{Non-recoverable work due to measurement}\label{appendix:Irrecoverable work due to measurement}

Recall that at the end of the measurement process, the average state of the compound system $\s+\aa$  is $\varrho\sub{\s+\aa}$ defined in \eq{eq:system-apparatus-after-measurement}, while the average state of $\s$  is $\rho_\tau := \tr\sub{\aa}[\varrho\sub{\s+\aa}] \equiv  \tr\sub{\aa}[U(\rho \otimes \xi)U^\dagger] \equiv \sum_{x \in \xx} \ii_x^\mm(\rho)$, and the average state of $\aa$ is $\xi_\tau := \tr\sub{\s}[\varrho\sub{\s+\aa}] \equiv \sum_{x\in \xx}p_\rho^M(x) \xi(x)$. Moreover, the average state of $\aa$ after premeasurement is $\xi' := \tr\sub{\s}[U(\rho \otimes \xi)U^\dagger]$.  Using the fact that $\sum_{x \in \xx} P\sub{\aa}^x = \one\sub{\aa}$, then by \eq{eq:conditional-work} the average work, over all measurement outcomes, can  be expressed as 
\begin{align}\label{eq:average-work}
\<\w_\rho^\mm(x)\> &:= \sum_{x \in \xx} p_\rho^M(x) \w_\rho^\mm(x), \nonumber \\
&= \tr[(H\sub{\s}(\tau) + H\sub{\aa}(\tau)) U(\rho\otimes \xi)U^\dagger] \nonumber \\
& \qquad - \tr[(H\sub{\s}(0) + H\sub{\aa}(0))(\rho\otimes \xi)] , \nonumber \\
&= \tr[H\sub{\s}(\tau)\rho_\tau] - \tr[H\sub{\s}(0)\rho] + \tr[H\sub{\aa}(\tau)\xi']  -\tr[H\sub{\aa}(0) \xi],\nonumber \\
& = \tr[H\sub{\s}(\tau)\rho_\tau] - \tr[H\sub{\s}(0)\rho] + \tr[H\sub{\aa}(\tau)\xi_\tau]  -\tr[H\sub{\aa}(0) \xi].
\end{align}
In the last line, we have used the fact that $Z\sub{\aa}$ commutes with $H\sub{\aa}(\tau)$ to infer that $\xi'$ and $\xi_\tau$ have the same expected energies. 

The free energy of a quantum state $\rho$, with respect to the Hamiltonian $H$ and  temperature $T$, is defined as 
\begin{align}\label{eq:free-energy}
F(\rho,H,T) := \tr[H \rho] - k_B T \, S(\rho) ,
\end{align}
where $k_B$ is Boltzmann's constant and $S(\rho) := -\tr[\rho \, \ln{\rho}]$ is the von Neumann entropy of $\rho$. Moreover, we note that since $\xi_\tau$ is a mixture of orthogonal states $\xi(x)$, it follows that $S(\xi_\tau) = \mathscr{H} + \sum_{x\in \xx} p_\rho^M(x) S(\xi(x))$, where $\mathscr{H}:= -\sum_{x\in\xx} p_\rho^M(x)\,  \ln{p_\rho^M(x)}$ is the Shannon entropy of the measurement probabilities. By using  \eq{eq:free-energy}, and the aforementioned relation, we can restate \eq{eq:average-work} as
\begin{align}\label{eq:average-work-free-energy-appendix}
\<\w_\rho^\mm(x)\> & =  \Delta F \sub{\s} +  \Delta F \sub{\aa} +  k_B T\big(S(\rho_\tau) + S(\xi_\tau) - S(\rho) - S(\xi)\big) \nonumber \\
& =  \Delta F \sub{\s} +  \Delta F \sub{\aa} + k_B T\big(I\sub{\s:\aa} + S(\xi_\tau) - S(\xi')\big), \nonumber \\
 &= \Delta F \sub{\s} +  \Delta F \sub{\aa} +  k_B T\big(I\sub{\s:\aa} + \mathscr{H}  -  \mathscr{X}\sub{\aa}\big).
\end{align}
Here $\Delta F\sub{\s} := F(\rho_\tau,H\sub{\s}(\tau),T) - F(\rho,H\sub{\s}(0),T)$  and $\Delta F\sub{\aa} := F(\xi_\tau,H\sub{\aa}(\tau),T) - F(\xi,H\sub{\aa}(0),T)$ are the increase in free energy of $\s$ and $\aa$, respectively; $I\sub{\s:\aa}:= S(\rho_\tau) + S(\xi') - S(U(\rho\otimes \xi)U^\dagger)$ is the quantum mutual information between $\s$ and $\aa$ after the premeasurement unitary evolution;    and $\mathscr{X}\sub{\aa}:= S(\xi') - \sum_{x\in \xx} p_\rho^M(x) S(\xi(x))$ is the Holevo information  of the apparatus with respect to the ideal measurement of $Z\sub{\aa}$ and state $\xi'$, which obeys the inequality $\mathscr{H} \geqslant \mathscr{X}\sub{\aa} \geqslant 0$ \cite{DenesPetz2008,Sagawa2012a}. Alternatively, the average work can be expressed as 
\begin{align}\label{eq:average-work-free-energy-appendix-2}
\<\w_\rho^\mm(x)\> & =  \Delta F \sub{\s+\aa} +  k_B T\big(S(\varrho\sub{\s+\aa}) - S(\rho\otimes\xi)\big) \nonumber \\
& =  \Delta F \sub{\s+\aa} + k_B T \big(I\sub{\s:\aa} - I\sub{\s:\aa}' + \mathscr{H}  -  \mathscr{X}\sub{\aa} \big),
\end{align}
where $ \Delta F \sub{\s+\aa} := F(\varrho\sub{\s+\aa}, H(\tau), T) - F(\rho\otimes \xi , H(0), T)$ is the increase in free energy of the compound system, where $H(t) = H\sub{\s}(t) + H\sub{\aa}(t)$ for $t=0, \tau$, and $I\sub{\s:\aa}':= S(\rho_\tau) + S(\xi_\tau) - S(\varrho\sub{\s+\aa})$ is the mutual information between $\s$ and $\aa$ at the end of the measurement process.

Returning the Hamiltonian of the compound system to $H\sub{\s}(0) + H\sub{\aa}(0)$, while returning the state of $\s$ and $\aa$ to $\rho$ and $\xi$ respectively, will incur a minimal work cost of  $-\Delta F \sub{\s} - \Delta F \sub{\aa}$ and $-\Delta F\sub{\s+\aa}$ if the systems are coupled to the thermal bath individually, or collectively, respectively.  In either case, the minimal work cost is achieved in the limit as the process is quasistatic \cite{Reeb2013a, Anders-thermo-discrete}. Therefore, the non-inclusive non-recoverable work, which does not make use of the correlations between $\s$ and $\aa$ during the closing of the cycle, can be defined as 
\begin{align}
\w_\mathrm{irr}^\mm(\rho) &:= \<\w_\rho^\mm(x)\>-\Delta F \sub{\s} - \Delta F \sub{\aa},\nonumber \\
& =   k_B T\big( I\sub{\s:\aa} +\mathscr{H}  -  \mathscr{X}\sub{\aa}\big),
\end{align}
while the inclusive non-recoverable work, which does make use of the correlations between $\s$ and $\aa$, is 
\begin{align}
\w_\mathrm{inc.irr}^\mm(\rho) &:= \<\w_\rho^\mm(x)\>-\Delta F \sub{\s+\aa},\nonumber \\
& =   k_B T\big( S(\varrho\sub{\s+\aa}) - S(\rho\otimes\xi)\big) \equiv \w_\mathrm{irr}^\mm(\rho) - k_B T \, I\sub{\s:\aa}'.
\end{align}

We note that the non-negativity of $\w_\mathrm{irr}^\mm(\rho)$ is guaranteed by the fact that $I\sub{\s:\aa} \geqslant 0$ and $\mathscr{H}  \geqslant   \mathscr{X}\sub{\aa} \geqslant 0$, while the non-negativity of $\w_\mathrm{inc. irr}^\mm(\rho)$ is guaranteed by the fact that the quantum channel that implements the transformation $\rho\otimes \xi \mapsto \varrho\sub{\s+\aa}$ is unital.  The non-negativity of the mutual information also implies that $\w_\mathrm{irr}^\mm(\rho)  \geqslant \w_\mathrm{inc.irr}^\mm(\rho) $.

\section{Non-recoverable work due to projective  measurements}\label{appendix:Non-recoverable work due to projective  measurements}

Consider the  self adjoint operator  $M = \sum_{x \in \xx} x M_x$ that defines a  projective-valued measure on $\h\sub{\s}$. Here, $M_x$ are projection operators of arbitrary rank.  We wish to implement this by the measurement model $\mm = (\h\sub{\aa}, \xi, U, Z\sub{\aa})$, where $\xi = \sum_{i=1}^r q_i \pra{\phi_i}$ (with $0<q_i <1$ and $\sum_i q_i =1$) is a mixed state of rank $1\leqslant r\leqslant \mathrm{dim}(\h\sub{\aa})$. By linearity, for each $i$,  $(\h\sub{\aa}, \pra{\phi_i}, U, Z\sub{\aa})$ must also be a measurement model for $M$ \cite{Busch2016a}. Denoting the eigenstates of $M$ as $\ket{\psi_x^\alpha}$, with $\alpha$ denoting the degeneracy,  a general (albeit not the most general) prescription of premeasurement will  be
\begin{align}\label{eq:Type-4-premeasurement}
U : \ket{\psi_x^\alpha} \otimes \ket{\phi_i} \mapsto  \ket{\tilde \psi_{x,i}^\alpha} \otimes \ket{\phi_{x,i}},
\end{align}
where  $\ket{\phi_{x,i}}$ is an eigenvector of $Z\sub{\aa}$ with eigenvalue $x$, while for each $i$ and $x$,  $\{\ket{\tilde \psi_{x,i}^\alpha}\}_\alpha$ is an orthonormal set of vectors that need not be eigenstates of $M$. This is referred to as a Type-4 measurement \cite{PeterMittelstaedt2004}. 

If upon observing outcome $x$ of $M$, a subsequent measurement of $M$ will yield $x$ with certainty, the measurement of $M$ is said to be repeatable. Given the premeasurement unitary defined in \eq{eq:Type-4-premeasurement}, the measurement will be repeatable if $\ket{\tilde \psi_{x,i}^\alpha} = V_{x,i} \ket{\psi_x^\alpha}$ are eigenstates of $M$ with eigenvalue $x$. Here, $V_{x,i}$ is a unitary operator that acts non-trivially only on the support of $M_x$.  Such a measurement model is referred to as a Type-3 measurement, and will implement   the instrument $\ii_x^\mathrm{Rep}$ defined as: 
\begin{align}
\ii_x^\mathrm{Rep}(\rho) &= \sum_{i=1}^r q_i V_{x,i} \ii_x^\mathrm{Ideal}(\rho) V_{x,i}^\dagger, \nonumber \\
\ii_x^\mathrm{Ideal}(\rho) & = M_x \rho M_x.
\end{align}
In the special case where, for each $i$, $V_{x,i} = \one$, then we have $\ii_x^\mathrm{Rep} = \ii_x^\mathrm{Ideal}$, which is a L\"uders instrument that implements an ideal (also known as a Type-2) measurement. If $M$ is a non-degenerate observable, then the only repeatable measurements are automatically ideal. These are the standard von Neumann measurements, also known as Type-1. 

If the measurement is repeatable,  it follows that for every $i\ne j$, there exists a pair of vectors $\ket{\tilde \psi_{x,i}^\alpha}$ and  $\ket{\tilde \psi_{x,j}^{\alpha'}}$ with a non-vanishing inner product.     As unitary operators preserve the inner product, this can only be achieved if $\<\phi_{x,i}| \phi_{x,j}\> = \delta_{i,j}$. Therefore,  the cardinality of the set of orthonormal vectors 
\begin{align}
\left\{\ket{\phi_{x,i}} : x \in \{1,\dots,|\xx|\}, i \in \{1,\dots,r\}\right \}
\end{align}
must be $r  \, |\xx|$, where $|\xx|$ is the cardinality of the outcome set $\xx$. Since this value cannot exceed the dimension of $\h\sub{\aa}$, the rank of $\xi$ must obey the inequality 
\begin{align}
r \leqslant \frac{\mathrm{dim}(\h\sub{\aa})}{|\xx|}.
\end{align}
In other words, $\xi$ cannot have full rank. 

Let us express the initial state of the system as $\rho = \sum_{x,y,\alpha,\beta} c_{x,y,\alpha,\beta}|\psi_x^\alpha\>\<\psi_y^\beta|$. Therefore, the conditional states of the compound system $\s+\aa$, after a repeatable measurement process, will be
\begin{align}\label{eq:sys-app-cond-state}
\varrho\sub{\s+\aa}^\mathrm{Rep}(x) &:= \frac{1}{p_\rho^M(x)}P\sub{\aa}^x U(\rho \otimes \xi)U^\dagger P\sub{\aa}^x, \nonumber \\
& = \sum_{i,x,y,\alpha,\beta} \frac{q_i  c_{x,y,\alpha,\beta}}{p_\rho^M(x)} |\tilde \psi_x^\alpha\>\< \tilde \psi_y^\beta| \otimes P\sub{\aa}^x|\phi_{x,i}\>\<\phi_{y,i}|P\sub{\aa}^x, \nonumber \\
& = \sum_{i,x,\alpha,\beta} \frac{q_i  c_{x,x,\alpha,\beta}}{p_\rho^M(x)} |\tilde \psi_x^\alpha\>\< \tilde \psi_x^\beta| \otimes \pra{\phi_{x,i}}, \nonumber \\
& = \sum_i q_i V_{x,i} \rho^\mathrm{Ideal}(x) V_{x,i}^\dagger \otimes \pra{\phi_{x,i}},
\end{align}
where $\rho^\mathrm{Ideal}(x):= \ii_x^\mathrm{Ideal}(\rho)/p_\rho^M(x)$. It is clear from this equation that the apparatus states representing outcome $x$ will be 
\begin{align}
\xi(x) := \tr\sub{\s}[\varrho\sub{\s+\aa}^\mathrm{Rep}(x) ]=  \sum_{i=1}^r q_i \pra{\phi_{x,i}}.
\end{align}
Therefore, the von Neumann entropy of $\xi(x)$ will be equal to that of $\xi$ for all $x\in \xx$. This implies that the Holevo information of the apparatus, with respect to the state $\xi'$ and ideal measurement of $Z\sub{\aa}$, is 
\begin{align}
\mathscr{X}\sub{\aa} & := S(\xi') - \sum_{x \in \xx}p_\rho^M(x) S(\xi(x)), \nonumber \\
& = S(\xi') - S(\xi).
\end{align} 
It follows that the non-inclusive non-recoverable work is, by \eq{eq:non-recoverable-work}, given as 
\begin{align}\label{eqapp:non-rec-work-rep}
\w_\mathrm{irr}^\mathrm{Rep}(\rho) &= k_B T\big(I\sub{\s:\aa}  + \mathscr{H} -  \mathscr{X}\sub{\aa} \big), \nonumber \\
& = k_B T \big( \mathscr{H} + S(\rho_\tau^\mathrm{Rep} ) - S(\rho) \big) \nonumber \\
& \quad+ k_B T \big(S(\xi')  - S(\xi)  -  \mathscr{X}\sub{\aa} \big), \nonumber \\
& = k_B T \big(\mathscr{H} + S(\rho_\tau^\mathrm{Rep} ) - S(\rho) \big),
\end{align}
where  $\rho^\mathrm{Rep}(x) := \ii_x^\mathrm{Rep}(\rho)/p_\rho^M(x)$ and $\rho_\tau^\mathrm{Rep} := \sum_{x\in \xx} \ii_x^\mathrm{Rep}(\rho)$.  Similarly, the inclusive non-recoverable work for repeatable measurements can be obtained from \eq{eq:inc-non-recoverable-work} to be
\begin{align}\label{eqapp:inc-non-rec-work-rep}
\w_\mathrm{inc.irr}^\mathrm{Rep}(\rho) &= k_B T\big(S(\varrho\sub{\s+\aa}^\mathrm{Rep}) - S(\rho\otimes \xi) \big), \nonumber \\
& = k_B T\left( \mathscr{H} + \sum_{x\in\xx} p_\rho^M(x) S(\varrho\sub{\s+\aa}^\mathrm{Rep}(x)) - S(\rho) - S(\xi)  \right), \nonumber \\
& = k_B T\left( \mathscr{H} + \sum_{x\in\xx} p_\rho^M(x)\big(S(\rho^\mathrm{Rep}(x)) + S(\xi(x)) - \ii\sub{\s:\aa}^\mathrm{Rep}(x)\big) - S(\rho) - S(\xi)  \right), \nonumber \\
& = k_B T\left( \mathscr{H} + \sum_{x\in\xx} p_\rho^M(x)\big( S(\rho^\mathrm{Rep}(x)) - \ii\sub{\s:\aa}^\mathrm{Rep}(x) \big) - S(\rho)  \right), \nonumber \\
& = k_B T \big( S(\rho_\tau^\mathrm{Ideal} ) - S(\rho) \big),
\end{align}
where $\ii\sub{\s:\aa}^\mathrm{Rep}(x):= S(\rho^\mathrm{Rep}(x)) + S(\xi(x)) - S(\varrho\sub{\s+\aa}^\mathrm{Rep}(x))$ is the quantum mutual information between $\s$ and $\aa$ conditional on outcome $x$, and $\rho_\tau^\mathrm{Ideal} := \sum_{x\in \xx} \ii_x^\mathrm{Ideal}(\rho)$. To see this, first note that because the states $\pra{\phi_{x,i}}$ are all orthogonal, the von Neumann entropy of the conditional states in  \eq{eq:sys-app-cond-state} will be given as 
\begin{align}
S(\varrho\sub{\s+\aa}^\mathrm{Rep}(x)) &= S(\xi) + \sum_i q_i S(V_{x,i} \rho^\mathrm{Ideal}(x) V_{x,i}^\dagger), \nonumber \\
& = S(\xi) + S(\rho^\mathrm{Ideal}(x)).
\end{align} 
Therefore, since $S(\xi(x)) = S(\xi)$, the conditional mutual information terms will be $\ii\sub{\s:\aa}^\mathrm{Rep}(x) = S(\rho^\mathrm{Rep}(x)) - S(\rho^\mathrm{Ideal}(x))$.
 
Note that  both $\w_\mathrm{irr}^\mathrm{Rep}(\rho)$ and $\w_\mathrm{inc.irr}^\mathrm{Rep}(\rho)$ do not depend on the apparatus specifics. However, while  $\w_\mathrm{irr}^\mathrm{Rep}(\rho)$ also depends on the particular instrument $\ii_x^\mathrm{Rep}$, the same does not hold for $\w_\mathrm{inc.irr}^\mathrm{Rep}(\rho)$, as the inclusive non-recoverable work will be the same for all  repeatable instruments.  Moreover,  it follows from Uhlmann's theorem \cite{Wehrl1978} that 
\begin{align}\label{eq:inequality-uhlmann}
\w_\mathrm{irr}^\mathrm{Rep}(\rho) - \w_\mathrm{irr}^\mathrm{Ideal}(\rho) &= k_B T \, \sum_{x\in \xx} p_\rho^M(x) \left[ S\left ( \sum_i q_i V_{x,i} \rho^\mathrm{Ideal}(x) V_{x,i}^\dagger  \right)  - S\left (\rho^\mathrm{Ideal}(x) \right) \right] \geqslant 0,\nonumber \\
\w_\mathrm{inc.irr}^\mathrm{Rep}(\rho) - \w_\mathrm{inc.irr}^\mathrm{Ideal}(\rho) & = 0
\end{align} 
with the equality condition in the first line being satisfied if and only if $V_{x,i} = V_x$ for all $i$ and $x$.

It has recently been suggested that, by the third law of thermodynamics, the preparation of the apparatus in a non-full rank state requires infinite resources and, as such, ideal (or repeatable) measurements are thermodynamically impossible \cite{Guryanova2018, Masanes2014}. To account for this, we consider measurement models where $\xi$ has full-rank, i.e. $r = \dim(\h\sub{\aa})$. We refer to such measurement models as  ``noisy'' projective measurements.

Although the smallest dimension permissible for the apparatus is $\dim(\h\sub{\aa}) = |\xx|$, wherin $\ket{ \phi_{x,i}}=\ket{\phi_x}$ for all $i$, we shall consider only the case where $\dim(\h\sub{\aa}) = \dim(\h\sub{\s})$. Here, $\dim(\h\sub{\aa}) \geqslant |\xx|$, with equality only when $M$ is a non-degenerate observable.  The simplest premeasurement unitary which will allow for a measurement of $M$ is a SWAP operator, followed by an appropriate local unitary on $\aa$.

We immediately see that here,  the mutual information terms in \eq{eq:non-recoverable-work} and \eq{eq:inc-non-recoverable-work} vanish, and so the inclusive and non-inclusive non-recoverable work will be identical. Moreover, since $S(\xi') = S(\rho)$, and $S(\xi(x)) = S(\rho^\mathrm{Ideal}(x))$,  we may identify $\mathscr{H} - \mathscr{X}\sub{\aa} = S(\rho_\tau^\mathrm{Ideal} ) - S(\rho)$, thereby obtaining the non-recoverable work
\begin{align}\label{eqapp:non-rec-work-noisy}
\w_\mathrm{irr}^\mathrm{Noisy}(\rho) &= \w_\mathrm{inc.irr}^\mathrm{Noisy}(\rho)=  k_B T\big(S(\rho_\tau^\mathrm{Ideal} ) - S(\rho)\big).
\end{align} 

Comparison between Eqs. \eqref{eqapp:non-rec-work-rep}, \eqref{eqapp:inc-non-rec-work-rep}, \eqref{eq:inequality-uhlmann} and \eqref{eqapp:non-rec-work-noisy} allows us to write the following relations: 

\begin{align}
\w_\mathrm{irr}^\mathrm{Rep}(\rho) &\geqslant \w_\mathrm{irr}^\mathrm{Ideal}(\rho) \geqslant  \w_\mathrm{irr}^\mathrm{Noisy}(\rho), \nonumber \\
\w_\mathrm{inc.irr}^\mathrm{Rep}(\rho) &= \w_\mathrm{inc.irr}^\mathrm{Ideal}(\rho) =  \w_\mathrm{inc.irr}^\mathrm{Noisy}(\rho).
\end{align}
Here, the first inequality in the top line results from \eq{eq:inequality-uhlmann}, while the second inequality is a consequence of $\w_\mathrm{irr}^\mathrm{Ideal}(\rho) - \w_\mathrm{irr}^\mathrm{Noisy}(\rho) = k_B T \, \mathscr{H}$.

\end{document}